%% file: mascots16.tex
\documentclass[conference]{IEEEtran}
\usepackage[tight,footnotesize]{subfigure}
\usepackage{booktabs, multicol, multirow}
\usepackage[acronym,nomain]{glossaries}
\usepackage[inline]{enumitem}
\usepackage{algorithm}
\usepackage{algorithmicx}
\usepackage{algpseudocode}
\usepackage{amsmath,amsfonts,amssymb}
\usepackage{subfigure}
\usepackage{grffile}
\usepackage{tabularx}
\usepackage{colortbl}
\usepackage{hhline}
\usepackage{hyperref}
\usepackage[flushleft]{threeparttable}
\usepackage{booktabs,caption,fixltx2e}
\usepackage{balance}
\newcommand{\bs}{\boldsymbol}
\renewcommand{\mathbf}[1]{{\boldsymbol #1}}
\hypersetup{
	colorlinks   = true,
	citecolor    = blue,
	linkcolor    = blue,
}
\usepackage[colorinlistoftodos,prependcaption,textsize=tiny]{todonotes}
\newcommand{\argmin}{\arg\!\min}
\newcommand{\argmax}{\arg\!\max}

\makeglossaries

\begin{document}
%

\title{An Uncertainty-Aware Approach to Optimal
Configuration of Stream Processing Systems}


\author{\IEEEauthorblockN{Pooyan Jamshidi}
	\IEEEauthorblockA{Imperial College London, UK\\
		Email: p.jamshidi@imperial.ac.uk}
	\and
	\IEEEauthorblockN{Giuliano Casale}
	\IEEEauthorblockA{Imperial College London, UK\\
		Email: g.casale@imperial.ac.uk}
}

\maketitle
\begin{abstract}

Finding optimal configurations for Stream Processing Systems (SPS) is a challenging problem due to the large number of parameters that can influence their performance and the lack of analytical models to anticipate the effect of a change. To tackle this issue, we consider tuning methods where an experimenter is given a limited budget of {experiments} and needs to carefully allocate this budget to find optimal configurations. We propose in this setting {B}ayesian {O}ptimization for {C}onfiguration {O}ptimization ({\small\sf BO4CO}), an auto-tuning algorithm that leverages Gaussian Processes (GPs) to iteratively capture posterior distributions of the configuration spaces and sequentially drive the experimentation. Validation based on Apache Storm demonstrates that our approach locates optimal configurations within a limited experimental budget, with an improvement of SPS performance typically of at least an order of magnitude compared to existing configuration algorithms.

\end{abstract}

%


\input{sections/introduction}

\input{sections/example}

\input{sections/methodology}

\input{sections/experiment}
\input{sections/discussion}
\input{sections/relatedwork}

\input{sections/conclusions}

\section*{Acknowledgment}
The research leading to these results has received funding from the European Commission as part of the DICE action (H2020-644869).

\bibliographystyle{abbrv}
{\footnotesize
	\bibliography{mascots16}}

\newpage
\appendix

\input{sections/appendix}

\end{document}

%% file: sections/introduction.tex
\section{Introduction}
\label{sec:introduction}

We live in an increasingly instrumented world, where a large number of heterogeneous data sources typically provide continuous data streams from live stock markets, video sources, production line status feeds, and vital body signals \cite{hirzel2014catalog}. Yet, the research literature lacks automated methods to support the configuration (\emph{i.e.}, auto-tuning) of the underpinning SPSs. One possible explanation is that, ``big data'' systems such as SPSs often combine emerging technologies that are still poorly understood from a performance standpoint \cite{johnstonperformance,yigitbasi2013towards} and therefore difficult to holistically configure. Hence there is a critical shortage of models and tools to anticipate the effects of changing a configuration in these systems. Examples of configuration parameters for a SPS include buffer size, heap sizes, serialization/de-serialization methods, among others.

Performance differences between a well-tuned configuration and a poorly configured one can be of orders of magnitude. Typically, administrators use a mix of rules-of-thumb, trial-and-error, and heuristic methods for setting configuration parameters. However, this way of testing and tuning is slow, and require skillful administrators with a good understanding of the SPS internals. Furthermore, decisions are also affected by the nonlinear interactions between configuration parameters. 

In this paper, we address the problem of finding optimal configurations under these requirements: (i) a configuration space composed by multiple parameters; (ii) a limited budget of experiments that can be allocated to test the system; (iii) experimental results affected by uncertainties due to measurement inaccuracies or intrinsic variability in the system processing times. While the literature on auto-tuning work is abundant with existing solutions for databases, e-commerce and batch processing systems that address some of the above challenges (\emph{e.g.}, rule-based  \cite{db2advisor}, design of experiment \cite{ustinova2015modelling}, model-based  \cite{menasce2001preserving,johnstonperformance,yigitbasi2013towards,Siegmund}, search-based \cite{xi2004smart,osogami2007optimizing,thonangi2008finding,hansen2001completely,behzad2013taming,ye2003recursive} and learning-based \cite{Bu2009}), this is the first work to consider the problem under such constraints altogether. 

In particular, we present a new auto-tuning algorithm called {\small\sf BO4CO} that leverages GPs \cite{gpml} to continuously estimate the mean and confidence interval of a response variable at yet-to-be-explored configurations. Using Bayesian optimization \cite{shahriaritaking}, the tuning process can account for all the available prior information about a system and the acquired configuration data, and apply a variety of kernel estimators \cite{Lizotte2012} to locate regions where optimal configuration may lie. To the best of our knowledge, this is the first time that GPs are used for automated system configuration, thus a goal of the present work is to introduce and apply this class of machine learning methods into system performance tuning.

{\small\sf BO4CO} is designed keeping in mind the limitations of sparse sampling from the configuration space. For example, its features include: (i) sequential planning to perform experiments that ensure coverage of the most promising zones; (ii) memorization of past-collected samples while planning new experiments; (iii) guarantees that optimal configurations will be eventually discovered by the algorithm. We show experimentally that {\small\sf BO4CO} outperforms previous algorithms for configuration optimization. Our real configuration datasets are collected for three different SPS benchmark systems, implemented with Apache Storm, and using 5 cloud clusters worth several months of experimental time.


The rest of this paper is organized as follows. Section \ref{sec:motivating-example} discusses the motivations. The {\small \sf BO4CO} algorithm is introduced in Section \ref{sec:method} and then validated in Section \ref{sec:experiments}. Finally, Section \ref{sec:discussion} discusses the applicability of {\small \sf BO4CO} in practice, Section \ref{sec:related-work} reviews state of the art and Section \ref{sec:conclusions} concludes the paper.

%% file: sections/example.tex
\section{Problem and Motivation}
\label{sec:motivating-example}

\subsection{Problem statement}
\label{sec:problem}
In this paper, we focus on the problem of optimal system configuration defined as follows. Let $X_i$ indicate the $i$-th configuration parameter, which takes values in a finite domain $Dom(X_i)$. In general, $X_i$ may either indicate (i) integer variable such as ``level of parallelism'' or (ii) categorical variable such as ``messaging frameworks'' or Boolean parameter such as ``enabling timeout''. Throughout the paper, by the term option, we mean possible values that can be assigned to a parameter. The \emph{configuration space} is thus $\mathbb{X}=Dom(X_1) \times \dots \times Dom(X_d)$, which is the Cartesian product of the domains of $d$ parameters of interest.
We assume that each configuration $\mathbf{x}\in \mathbb{X}$ is valid and denote by $f(\mathbf{x})$ the response measured on the SPS under that configuration. Throughout, we assume that $f$ is latency, however other response metrics (\emph{e.g.}, throughput) may be used. The graph of $f$ over configurations is called the \emph{response surface} and it is partially observable, \emph{i.e.,} the actual value of $f(\mathbf{x})$ is known only at points $\mathbf{x}$ that has been previously experimented with. 
We here consider the problem of finding an optimal configuration $\mathbf{x}^*$ that minimizes $f$ over the configuration space $\mathbb{X}$ with as few experiments as possible:
\begin{equation} \label{eq:reponse-function}
\mathbf{x}^*=\arg \min _{\mathbf{x}\in\mathbb{X}} f(\mathbf{x})
\end{equation}
In fact, the response function $f(\cdot)$ is usually unknown or partially known, \emph{i.e.}, $y_i=f(\mathbf{x}_i), \mathbf{x}_i \subset \mathbb{X}$. In practice, such measurements may contain noise, \emph{i.e.}, $y_i=f(\mathbf{x}_i)+\epsilon_i$.
Note that since the response surface is only partially-known, finding the optimal configuration is a blackbox optimization problem \cite{Lizotte2012,Saboori2008}, which is also subject to noise. In fact, the problem of finding a optimal solution of a non-convex and multi-modal response surface (cf. Figure \ref{fig:response-latency-wc-motivation}) is $\mathcal{NP}$-hard \cite{weise2009global}. Therefore, on instances where it is impossible to locate a global optimum, {\small \sf BO4CO} will strive to find the best possible local optimum within the available experimental budget.

\subsection{Motivation}
\label{sec:motivation}

\subsubsection{A running example}
\label{sec:example}
\begin{figure}[t]
	\begin{center}
		\includegraphics[width=7cm]{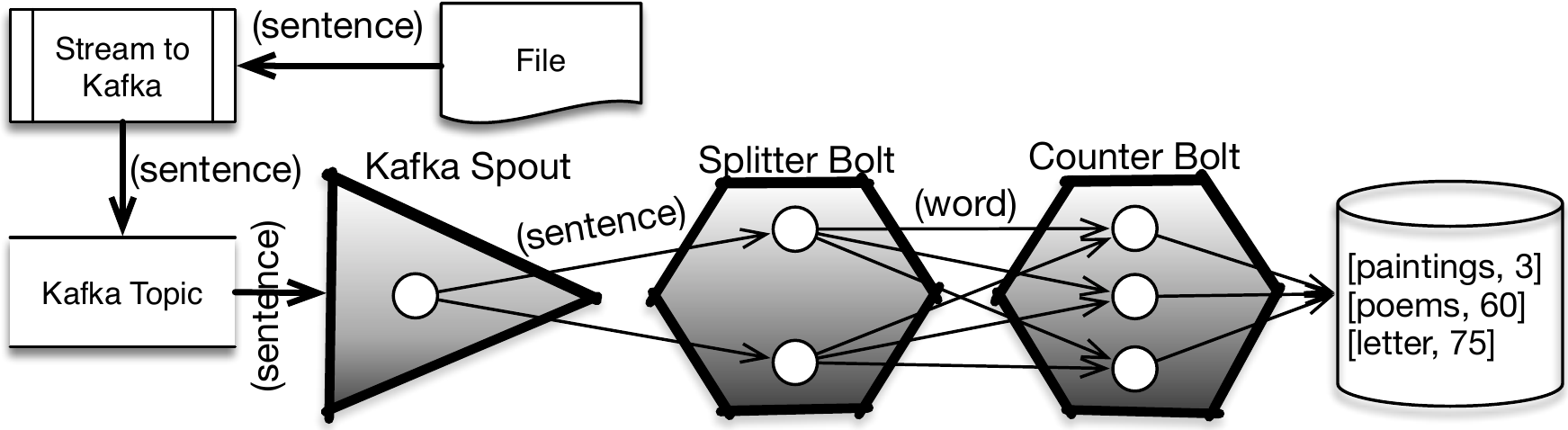}
		\caption{ {\sf WordCount}  topology architecture.}
		\label{fig:wc-arch}
	\end{center}
\end{figure}

{\sf WordCount} (cf. Figure \ref{fig:wc-arch}) is a popular benchmark SPS. In {\sf WordCount} a text file is fed to the system and it counts the number of occurrences of the words in the text file. In Storm, this corresponds to the following operations. A Processing Element (PE) called Spout is responsible to read the input messages (tuples) from a data source (\emph{e.g.,} a Kafka topic) and stream the messages (\emph{i.e.}, sentences) to the topology. Another PE of type Bolt called Splitter is responsible for splitting sentences into words, which are then counted by another PE called Counter. 

\subsubsection{Nonlinear interactions}
\label{sec:interaction}
\begin{figure}[t]
	\begin{center}
		\includegraphics[width=7cm]{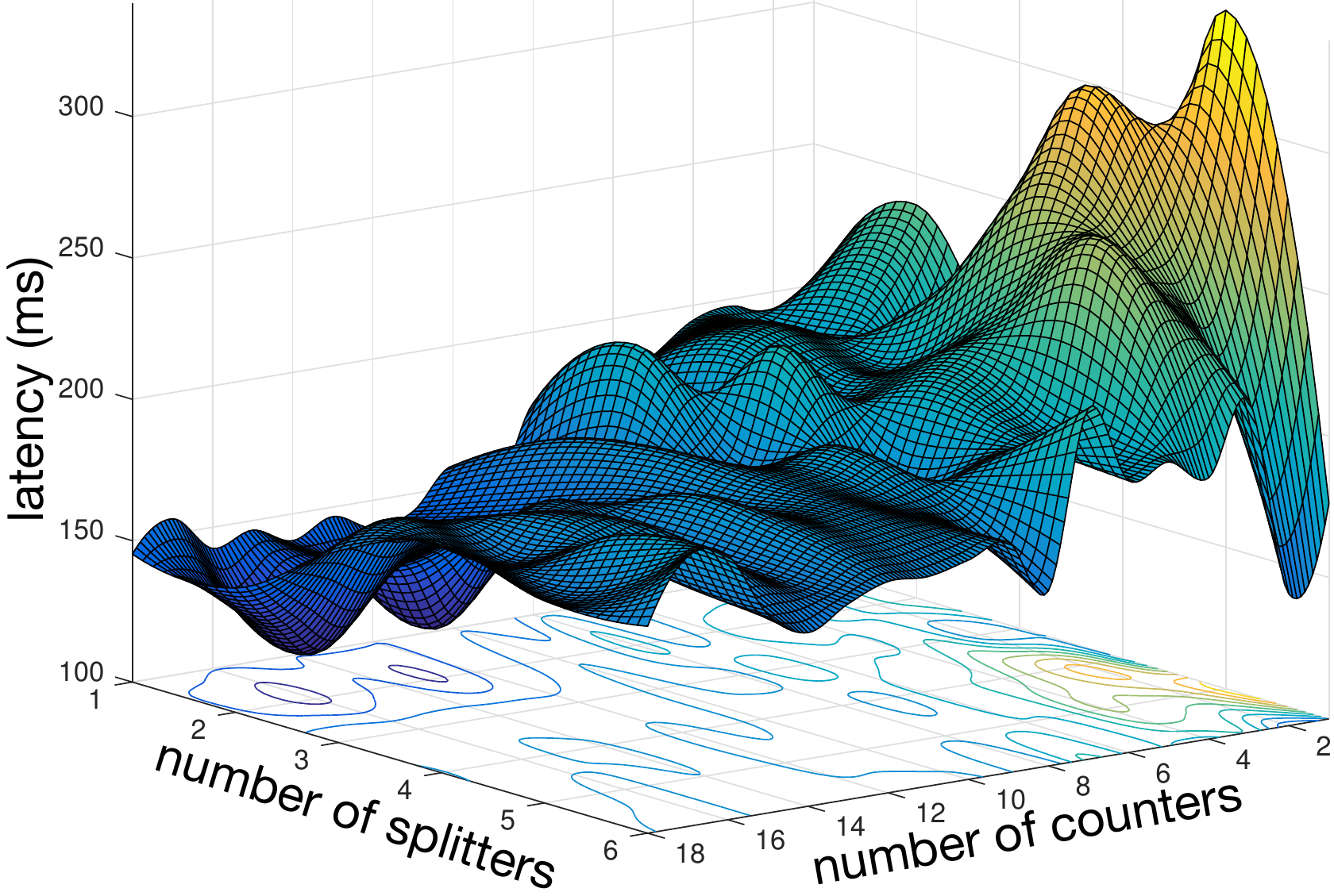}
		\caption{{\sf WordCount} response surface. It is an interpolated surface and is a projection of 6 dimensions, in {\sf wc(6D)}, onto 2D. It shows the non-convexity, multi-modality and the substantial performance difference between different configurations.}
		\label{fig:response-latency-wc-motivation}
	\end{center}
\end{figure}

\begin{figure}[t]
	\begin{center}
		\includegraphics[width=5cm]{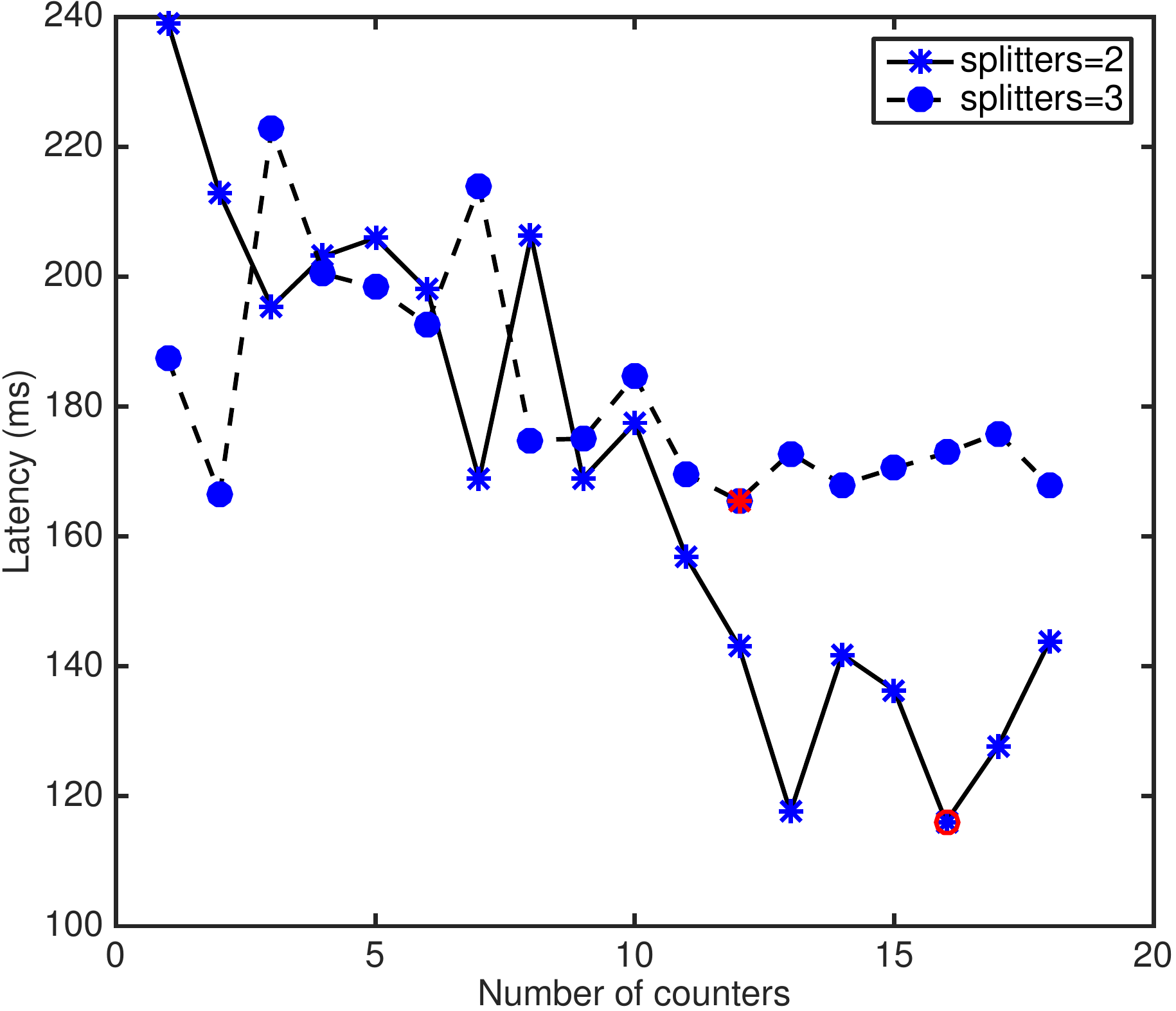}
		\caption{{\sf WordCount} latency, cut though Figure \ref{fig:response-latency-wc-motivation}.}
		\label{fig:response-latency-wc-splitters23}
	\end{center}
\end{figure}

We now illustrate one of the inherent challenges of configuration optimization. The metric that defines the surface in Figure~\ref{fig:response-latency-wc-motivation} is the {\em latency} of individual messages, defined as the time since emission by the Kafka Spout to completion at the Counter, see Figure \ref{fig:wc-arch}. Note that this function is the subset of {\sf wc(6D)} in Table \ref{tab:configuration-parameters} when the level of parallelism of Splitters and Counters is varied in $[1,6]$ and $[1, 18]$.
The surface is strongly \emph{non-linear} and \emph{multi-modal} and indicates two important facts. First, the performance difference between the best and worst settings is substantial, $65\%$, and with more intense workloads we have observed differences in latency as large as $99\%$, see Table \ref{tab:performance-gain}. Next, non-linear relations between the parameters imply that the optimal number of counters depends on the number of Splitters, and vice-versa.
Figure \ref{fig:response-latency-wc-splitters23} shows this \emph{non-linear interaction} \cite{Siegmund} and demonstrates that if one tries to minimize latency by acting just on one of these parameters at the time, the resulting configuration may not lead to a global optimum, as the number of Splitters has a strong influence on the optimal counters. 

\subsubsection{Sparsity of effects}
\label{sec:parameters-interactions}
Another observation from our extensive experiments with SPS is the \emph{sparsity of effects}. More specifically, this means low-order interactions among a few dominating factors can explain the main changes in the response function observed in the experiments. In this work we assume sparsity of effects, which also helps in addressing the intractable growth of the configuration space \cite{Kleijnen2010}. 

{\em Methodology}. In order to verify to what degree the sparsity of effects assumption holds in SPS, we ran experiments on 3 different benchmarks that exhibit different bottlenecks: {\sf WordCount} ({\sf wc}) is CPU intensive, {\sf RollingSort} ({\sf rs}) is memory intensive, and {\sf SOL} ({\sf sol}) is network intensive. Different testbed settings were also considered, for a total of 5 datasets, as listed in Table \ref{tab:configuration-parameters}. Note that the parameters we consider here are known to significantly influence latency, as they have been chosen according to professional tuning guides~\cite{nabi2014streams} and also small scale tests where we varied a single parameter to make sure that the selected parameters were all influential. For each test in the experiment, we run the benchmark for 8 minutes including the initial burn-in period. 
Further details on the experimental procedure are given in Section \ref{sec:experiment-design}. Note that the largest dataset (\emph{i.e.,} {\sf rs(6D)}) has required alone $3840\times8/60/24=21$ days, within a total experimental time of about 2.5 months to collect the datasets of Table \ref{tab:configuration-parameters}. 

{\em Results}. After collecting experimental data, we have used a common correlation-based feature selector\footnote{The most significant parameters are selected based on the following merit function \cite{Hall1999}, also shown in Table \ref{tab:configuration-parameters}:
\begin{equation} \label{eq:merit}
m_{ps}=\frac{n\overline{r_{lp}}}{\sqrt{n+n(n-1)\overline{r_{pp}}}},
\end{equation}
where $\overline{r_{lp}}$ is the mean parameter-latency correlation, $n$ is the number of parameters, $\overline{r_{pp}}$ is the average feature-feature inter-correlation \cite[Sec 4.4]{Hall1999}.} implemented in Weka to rank parameter subsets according to a heuristic. The bias of the merit function is toward subsets that contain parameters that are highly correlated with the response variable. Less influential parameters are filtered because they will have low correlation with latency, and a set with the main factors is returned. 
For all of the 5 datasets, we list in Table \ref{tab:configuration-parameters} the main factors. The analysis results demonstrate that in all the 5 experiments at most 2-3 parameters were strongly interacting with each other, out of a maximum of 6 parameters varied simultaneously. Therefore, the determination of the regions where performance is optimal will likely be controlled by such dominant factors, even though the determination of a global optimum will still depends on all the parameters.

\newcolumntype{g}{>{\columncolor{blue!25}}c}

\begin{table}[]
	\centering
	\caption{\small Sparsity of effects on 5 experiments where we have varied different subsets of parameters and used different testbeds. Note that these are the datasets we experimentally measured on the benchmark systems and we use them for the evaluation, more details including the results for 6 more experiments are in the appendix.} 
	\label{tab:configuration-parameters}
	\resizebox{\columnwidth}{!}{%
		\begin{threeparttable}
			\begin{tabular}{@{}lllccgc@{}}
				\toprule
				&\textbf{Topol.} & \multicolumn{1}{c}{\textbf{Parameters}}                                                                                      & \multicolumn{1}{c}{\textbf{Main factors}} & \multicolumn{1}{c}{\bf Merit} & \multicolumn{1}{c}{\textbf{Size}} & \multicolumn{1}{c}{\textbf{Testbed}}                                                                             \\ \midrule
				1  & {\sf wc(6D) }               & \begin{tabular}[c]{@{}l@{}}{\sf 1-spouts, 2-max\_spout, }\\ {\sf 3-spout\_wait, 4-splitters,}\\ {\sf 5-counters, 6-netty\_min\_wait} \end{tabular}         & \{1, 2, 5\} & 0.787                           & 2880 & C1                 \\ \midrule
				2  & {\sf sol(6D) }             & \begin{tabular}[c]{@{}l@{}}{\sf 1-spouts, 2-max\_spout,} \\ {\sf 3-top\_level, 4-netty\_min\_wait,} \\ {\sf 5-message\_size, 6-bolts} \end{tabular} & \{1, 2, 3\}& 0.447                           & 2866 & C2                          \\ \midrule
				3  & {\sf rs(6D) }           & \begin{tabular}[c]{@{}l@{}}{\sf 1-spouts, 2-max\_spout, }\\ {\sf 3-sorters, 4-emit\_freq,}\\ {\sf 5-chunk\_size, 6-message\_size} \end{tabular}          & \{3\}& 0.385                                 & 3840 & C3              \\ \midrule
				4  & {\sf wc(3D)}            & \begin{tabular}[c]{@{}l@{}}{\sf 1-max\_spout, 2-splitters,} \\ {\sf 3-counters} \end{tabular}                                                    & \{1, 2\}& 0.480                              & 756 & C4                            \\ \midrule
				5  & {\sf wc(5D)}            & \begin{tabular}[c]{@{}l@{}}{\sf 1-spouts, 2-splitters,} \\ {\sf 3-counters,} \\ {\sf 4-buffer-size, 5-heap} \end{tabular}                                         & \{1\}& 0.851                                 & 1080 & C5                            \\ \midrule
			\end{tabular}
		\end{threeparttable}}
	\end{table}
	
	 \begin{figure}[t]
	 	\begin{center}
	 		\includegraphics[width=6cm]{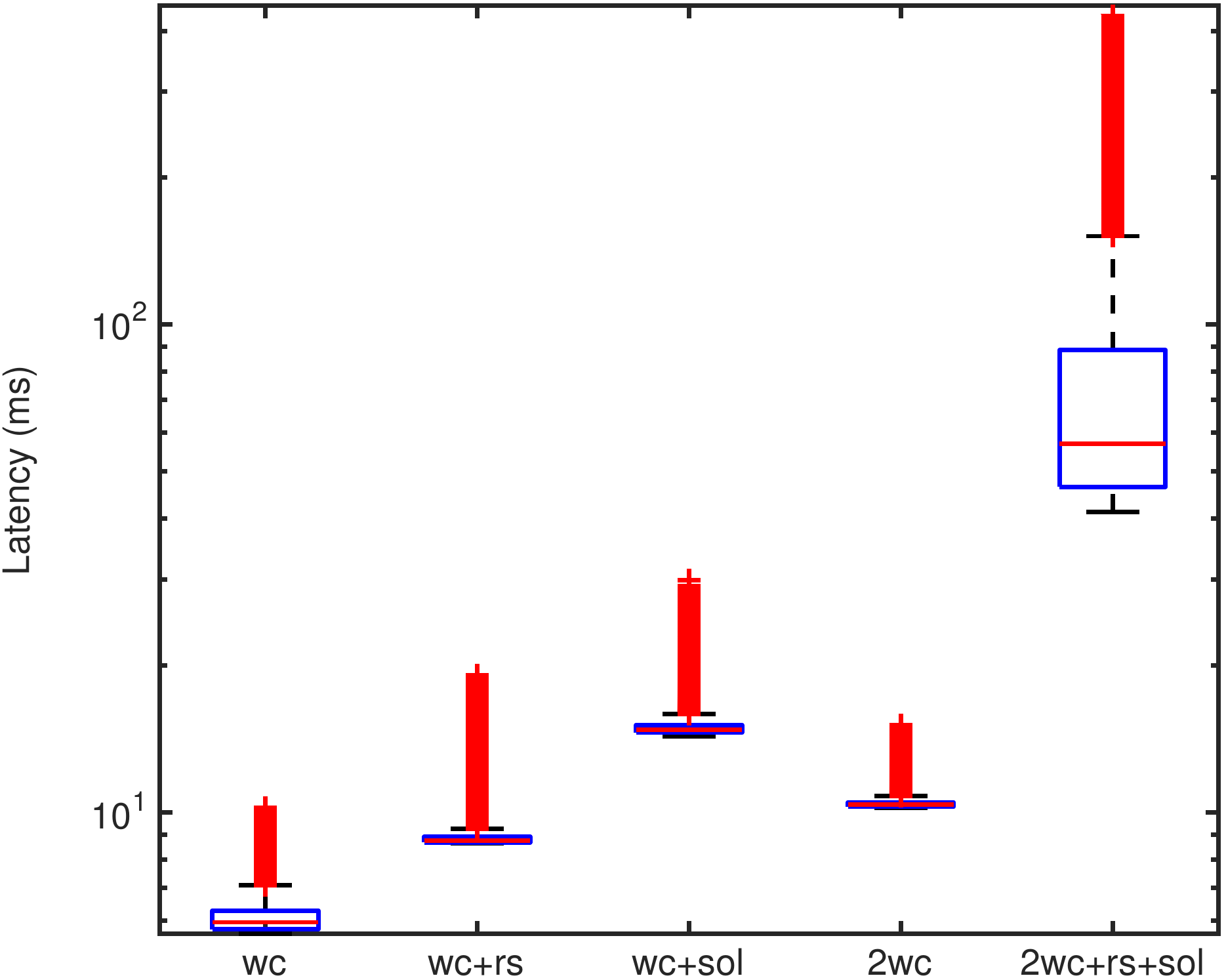}
	 		\caption{Noisy experimental measurements. Note that $+$ here means that {\sf wc} is deployed in a multi-tenant environment with other topologies and as a result not only the latency is increased but also the variability became greater.}
	 		\label{fig:boxplot-longrun}
	 	\end{center}
	 \end{figure}
	
\subsubsection{Measurement uncertainty}
\label{sec:uncertain-measurement}
We now illustrate measurement variabilities, which represent an additional challenge for configuration optimization. As depicted in Figure~\ref{fig:boxplot-longrun}, we took different samples of the latency metric over 2 hours for five different deployments of {\sf WordCount}. The experiments run on a multi-node cluster on the EC2 cloud. After filtering the initial burn-in, we computed averages and standard deviation of the latencies. 
Note that the configuration across all 5 settings is similar, the only difference is the number of co-located topologies in the testbed. 
The data in boxplots illustrate that variability can be small in some settings (\emph{e.g.,} {\sf wc}), while they can be large in some other experimental setups (\emph{e.g.,} {\sf 2wc+rs+sol}). In traditional techniques such as design of experiments, such variability is addressed by repeating experiments multiple times and obtaining regression estimates for the system model across such repetitions. However, we here pursue the alternative approach of relying on GP models to capture both mean and variance of measurements within the model that guides the configuration process. The theory underpinning this approach is discussed in the next section.

%% file: sections/methodology.tex
\section{BO4CO: Bayesian Optimization for Configuration Optimization}
\label{sec:method}


\subsection{Bayesian Optimization with Gaussian Process prior}

Bayesian optimization is a sequential design strategy that allows us to perform global optimization of blackbox functions \cite{shahriaritaking}. The main idea of this method is to treat the blackbox objective function $f(\bs x)$ as a random variable with a given prior distribution, and then perform optimization on the posterior distribution of $f(\bs x)$ given experimental data. In this work, GPs are used to model this blackbox objective function at each point $\bs x\in \mathbb{X}$. That is, let $\mathbb{S}_{1:t}$ be the experimental data collected in the first $t$ iterations and let $\mathbf{x}_{t+1}$ be a candidate configuration that we may select to run the next experiment. Then {\small\sf BO4CO} assesses the probability that this new experiment could find an optimal configuration using the posterior distribution:
\[
\Pr(f_{t+1}|\mathbb{S}_{1:t},\mathbf{x}_{t+1})\sim \mathcal{N}(\mu_{t}(\mathbf{x_{t+1}}),\sigma_t^2(\mathbf{x}_{t+1})),
\]
where $\mu_{t}(\mathbf{x_{t+1}})$ and $\sigma_t^2(\mathbf{x}_{t+1})$ are suitable estimators of the mean and standard deviation of a normal distribution that is used to model this posterior. The main motivation behind the choice of GPs as prior here is that it offers a framework in which reasoning can be not only based on mean estimates but also the variance, providing more informative decision makings. The other reason is that all the computations in this framework are based on \emph{linear algebra}.

\begin{figure}[t]
	\begin{center}
		\includegraphics[width=6cm]{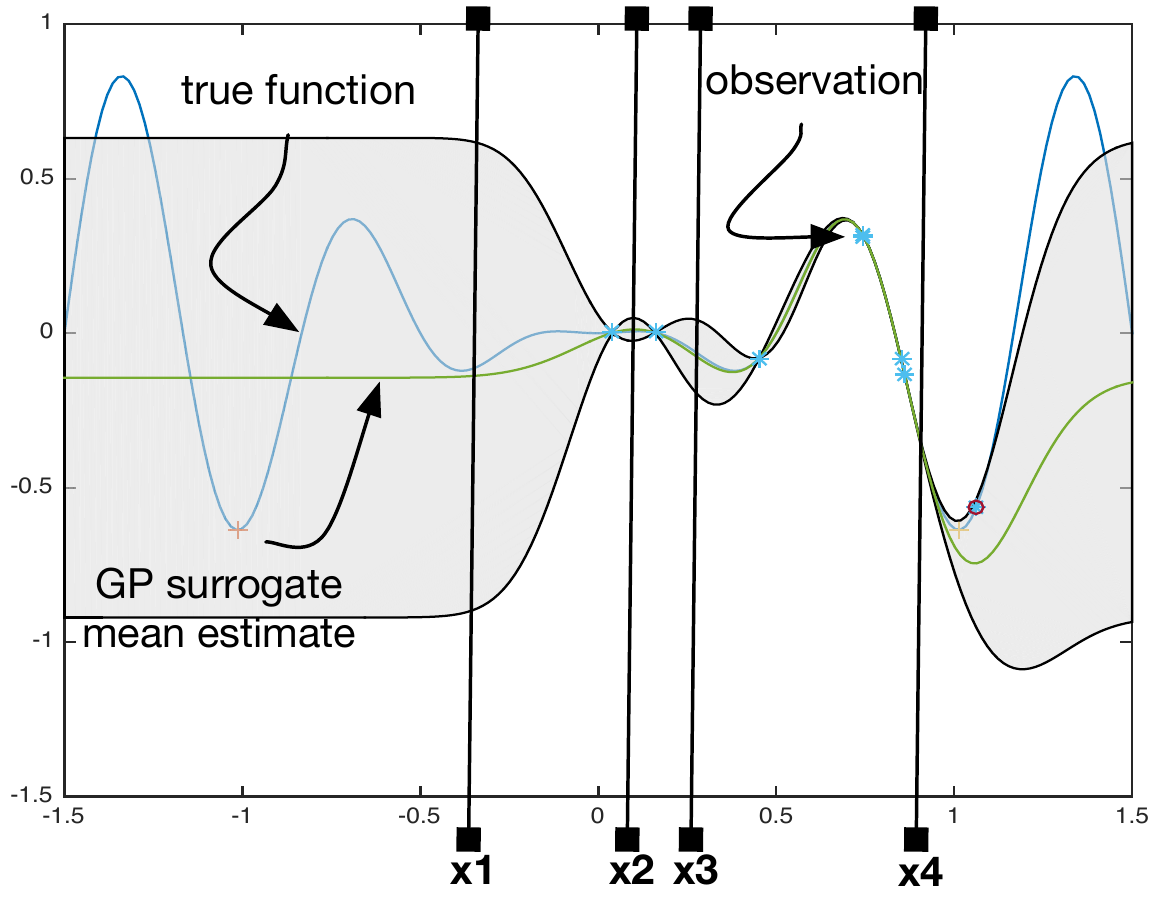}
		\caption{An example of 1D GP model: GPs provide mean estimates as well as the uncertainty in estimations, \emph{i.e.}, variance. } 
		\label{fig:example-gp}
	\end{center}
\end{figure}

Figure \ref{fig:example-gp} illustrates the GP-based Bayesian optimization using a 1-dimensional response surface. The curve in blue is the unknown true posterior distribution, whereas the mean is shown in green and the 95\% confidence interval at each point in the shaded area. Stars indicate measurements carried out in the past and recorded in $\mathbb{S}_{1:t}$ (\emph{i.e.,} observations). Configuration corresponds to $\mathbf{x}_1$ has a large confidence interval due to lack of observations in its neighborhood. Conversely, $\mathbf{x}_4$ has a narrow confidence since neighboring configurations have been experimented with. The confidence interval in the neighborhood of $\mathbf{x}_2$ and $\mathbf{x}_3$ is not high and correctly our approach does not decide to explore these zones. The next configuration $\mathbf{x}_{t+1}$, indicated by a small circle right to the $\mathbf{x}_4$, is selected based on a criterion that will be defined later.

%
%

\begin{figure}[t]
	\begin{center}
		\includegraphics[width=8.5cm]{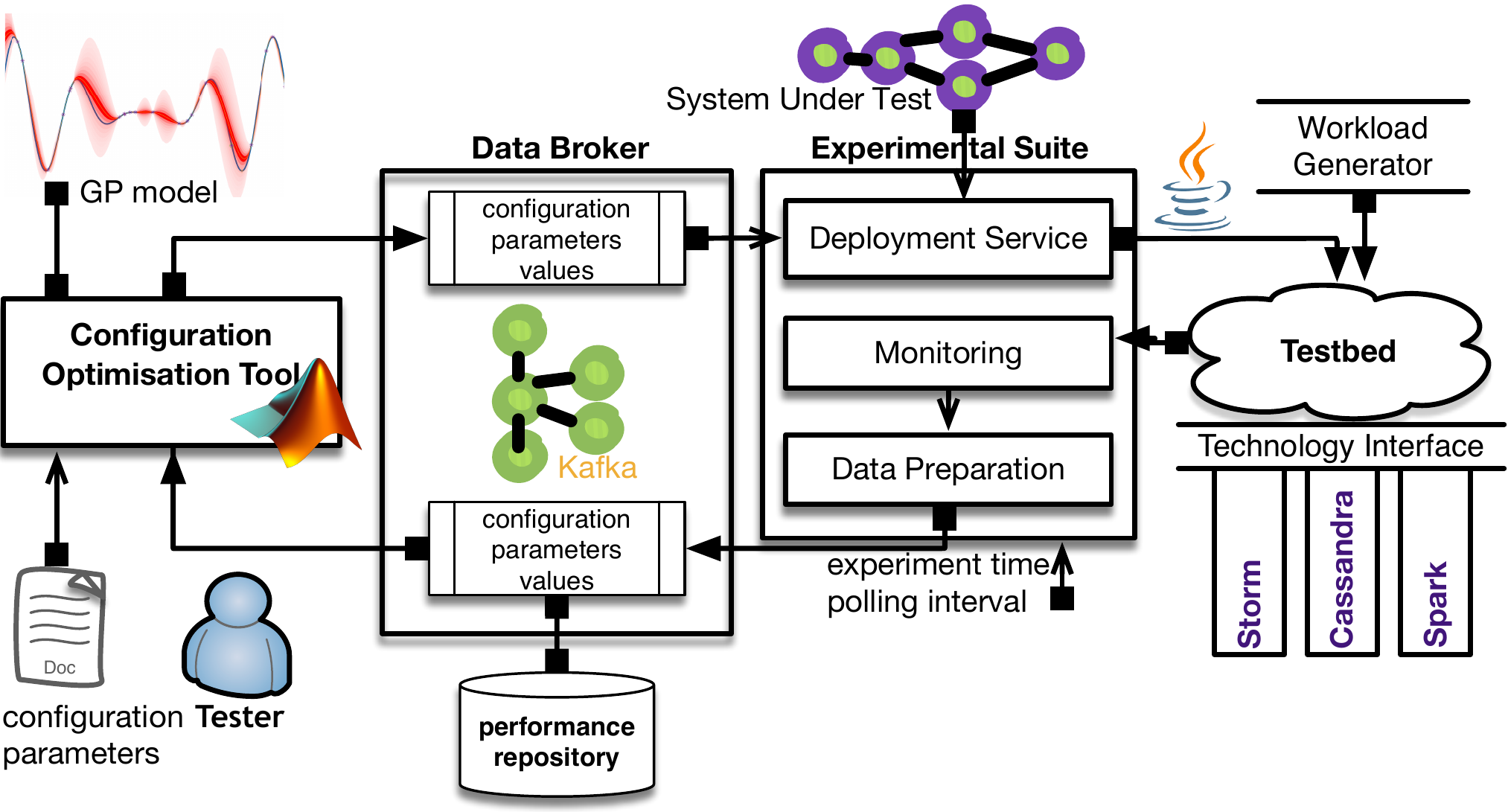}
		\caption{{\sf BO4CO} architecture: (i) optimization and (ii) experimental suite are integrated via (iii) a data broker. The integrated solution is available: \url{https://github.com/dice-project/DICE-Configuration-BO4CO}.}
		\label{fig:configuration-optimizer}
	\end{center}
\end{figure}

\algnewcommand{\LineComment}[1]{\(\triangleright\) #1}
\begin{tiny}
	\begin{algorithm}[t]
		\caption{: \textbf{BO4CO}}
		\label{alg:bo-alg}
		\begin{algorithmic}[1]
			\Require{Configuration space $\mathbb{X}$, Maximum budget $N_{max}$, Response function $f$, Kernel function $\mathbf{K}_\theta$, Hyper-parameters $\boldsymbol{\theta}$, Design sample size $n$, learning cycle $N_l$}
			\Ensure{Optimal configurations $\mathbf{x}^*$ and learned model $\mathcal{M}$}
			\State choose an initial sparse design (\emph{lhd}) to find an initial design samples $\mathcal{D}=\{\mathbf{x}_1, \dots ,\mathbf{x}_n\}$
			\State obtain \emph{performance measurements} of the initial design, $y_i \leftarrow f(\mathbf{x}_i)+\epsilon_i,\forall\mathbf{x}_i\in\mathcal{D}$
			\State $\mathbb{S}_{1:n} \leftarrow \{(\mathbf{x}_{i},y_{i})\}_{i=1}^n; t \leftarrow n+1$ 
			\State $\mathcal{M}(\mathbf{x}|\mathbb{S}_{1:n},\boldsymbol{\theta})\leftarrow$ fit a $\mathcal{GP}$ model to the design \Comment{Eq.\eqref{eq:GP draw}} 
			\While {$t \le N_{max}$}
			\State if $(t \mod N_l=0) \quad \boldsymbol{\theta} \leftarrow$ \emph{learn} the kernel hyper-parameters by maximizing the likelihood 
			\State find \emph{next configuration} $\mathbf{x}_t$ by optimizing the selection criteria over the estimated response surface given the data, $\mathbf{x}_t \leftarrow \arg max_\mathbf{x} u(\mathbf{x}| \mathcal{M},\mathbb{S}_{1:t-1})$ \Comment{Eq.\eqref{eq:next-point}} 
			\State obtain performance for the \emph{new configuration} $\mathbf{x}_t$, $y_t \leftarrow f(\mathbf{x}_t)+\epsilon_t$
			\State Augment the configuration  $\mathbb{S}_{1:t}=\{\mathbb{S}_{1:t-1},(\mathbf{x}_{t},y_{t})\}$
			\State $\mathcal{M}(\mathbf{x}|\mathbb{S}_{1:t},\boldsymbol{\theta})\leftarrow$ \textit{re-fit} a new GP model \Comment{Eq.\eqref{eq:gp-surrogate-mean-sigma}} 
			\State $t \leftarrow t+1$
			\EndWhile
			\State $(\mathbf{x}^*,y^*)=\min \mathbb{S}_{1:N_{max}}$
			\State  $\mathcal{M}(\mathbf{x})$
		\end{algorithmic}
		\vspace{2mm}
	\end{algorithm}
\end{tiny}

\subsection{BO4CO algorithm}



{\small\sf BO4CO}'s high-level architecture is shown in Figure \ref{fig:configuration-optimizer} and the procedure that drives the optimization is described in Algorithm. 
We start by bootstrapping the optimization following Latin Hypercube Design (lhd) to produce an initial design $\mathcal{D} = \{\mathbf{x}_1, \dots ,\mathbf{x}_n\}$ (cf. \emph{step 1} in Algorithm \ref{alg:bo-alg}). Although other design approaches (\emph{e.g.}, random) could be used, we have chosen lhd because: (i) it ensures that the configuration samples in $\mathcal{D}$ is representative of the configuration space $\mathbb{X}$, whereas traditional random sampling \cite{liebig2013scalable,henard2015combining} (called brute-force) does not guarantee this \cite{montgomery2008design}; (ii) another advantage is that the lhd samples can be taken one at a time, making it efficient in high dimensional spaces. 
After obtaining the measurements regarding the initial design, {\small \sf BO4CO} then fits a GP model to the design points $\mathcal{D}$ to form our belief about the underlying response function (cf. \emph{step 3} in Algorithm \ref{alg:bo-alg}). The while loop in Algorithm \ref{alg:bo-alg} iteratively updates the belief until the budget runs out: As we accumulate the data $\mathbb{S}_{1:t}=\{(\mathbf{x}_{i},y_{i})\}_{i=1}^t$, where $y_i=f(\mathbf{x}_i)+\epsilon_i$ with $\epsilon \sim \mathcal{N}(0,\sigma^2)$, a prior distribution $\Pr(f)$ and the likelihood function $\Pr(\mathbb{S}_{1:t}|f)$ form the posterior distribution: $\Pr(f|\mathbb{S}_{1:t}) \propto \Pr(\mathbb{S}_{1:t}|f)\Pr(f)$. 

A GP is a distribution over functions \cite{gpml}, specified by its mean (see Section \ref{sec:prior}), and covariance (see Section \ref{sec:kernel}):
\begin{equation} \label{eq:GP draw}
y=f(\mathbf{x}) \sim \mathcal{GP} (\mu(\mathbf{x}), k(\mathbf{x}, \mathbf{x}')),
\end{equation}
where $k(\mathbf{x}, \mathbf{x}')$ defines the distance between $\mathbf{x}$ and $\mathbf{x}'$. 
Let us assume $\mathbb{S}_{1:t}=\{(\mathbf{x}_{1:t},y_{1:t}) | y_i:=f(\mathbf{x}_i)\}$ be the collection of $t$ observations. 
The function values are drawn from a multi-variate Gaussian distribution $\mathcal{N}(\boldsymbol{\mu},\mathbf{K})$, where $\mathbf{\mu}:=\mu(\mathbf{x}_{1:t})$,
\begin{equation} \label{eq:covariance}
\mathbf{K}:=
\begin{bmatrix}
k(\mathbf{x}_1,\mathbf{x}_1)  &  \dots & k(\mathbf{x}_1,\mathbf{x}_t)   \\
\vdots  & \ddots &  \vdots \\
k(\mathbf{x}_t,\mathbf{x}_1)  &  \dots & k(\mathbf{x}_t,\mathbf{x}_t) 
\end{bmatrix}
\end{equation}

In the while loop in {\small \sf BO4CO}, given the observations we accumulated so far, we intend to fit a new GP model: 
\begin{equation} \label{eq:joint-gp}
\begin{bmatrix}
\mathbf{f}_{1:t} \\
f_{t+1}
\end{bmatrix}
\sim \mathcal{N}(\boldsymbol{\mu},
\begin{bmatrix}
\mathbf{K}+\sigma^2\mathbf{I} & \mathbf{k} \\
\mathbf{k}^\intercal & k(\mathbf{x}_{t+1},\mathbf{x}_{t+1}) 
\end{bmatrix}),
\end{equation}
where $\mathbf{k}(\mathbf{x})^\intercal=[k(\mathbf{x},\mathbf{x}_1) \quad k(\mathbf{x},\mathbf{x}_2) \quad \dots \quad k(\mathbf{x},\mathbf{x}_t)]$ and $\mathbf{I}$ is identity matrix.
Given the Eq. \eqref{eq:joint-gp}, the new GP model can be drawn from this new Gaussian distribution:
\begin{equation} \label{eq:gp-surrogate}
\begin{aligned}
\Pr(f_{t+1}|\mathbb{S}_{1:t},\mathbf{x}_{t+1})=\mathcal{N}(\mu_{t}(\mathbf{x_{t+1}}),\sigma_t^2(\mathbf{x}_{t+1})),
\end{aligned}
\end{equation}
where
\begin{align}
\label{eq:gp-surrogate-mean-sigma}
\mu_{t}(\mathbf{x})&=\mu(\mathbf{x})+\mathbf{k}(\mathbf{x})^\intercal (\mathbf{K}+\sigma^2\mathbf{I})^{-1} (\mathbf{y}-\boldsymbol{\mu}) \\
\sigma_t^2(\mathbf{x})&=k(\mathbf{x},\mathbf{x})+\sigma^2\mathbf{I} - \mathbf{k}(\mathbf{x})^\intercal (\mathbf{K}+\sigma^2\mathbf{I})^{-1} \mathbf{k}(\mathbf{x})
\end{align}
These posterior functions are used to select the next point $\mathbf{x}_{t+1}$ as detailed in Section \ref{sec:acquision-functions}. 

\subsection{Configuration selection criteria}
\label{sec:acquision-functions}

The selection criteria is defined as $u:\mathbb{X}\rightarrow\mathbb{R}$ that selects $\mathbf{x}_{t+1}\in\mathbb{X}$, should $f(\cdot)$ be evaluated next (\emph{step 7}):
\begin{equation} \label{eq:next-point}
\mathbf{x}_{t+1}=\argmax_{\mathbf{x}\in \mathbb{X}} u(\mathbf{x}|\mathcal{M},\mathbb{S}_{1:t})
\end{equation}

%
%
%
%

\begin{figure}[t]
	\begin{center}
		\includegraphics[width=5cm]{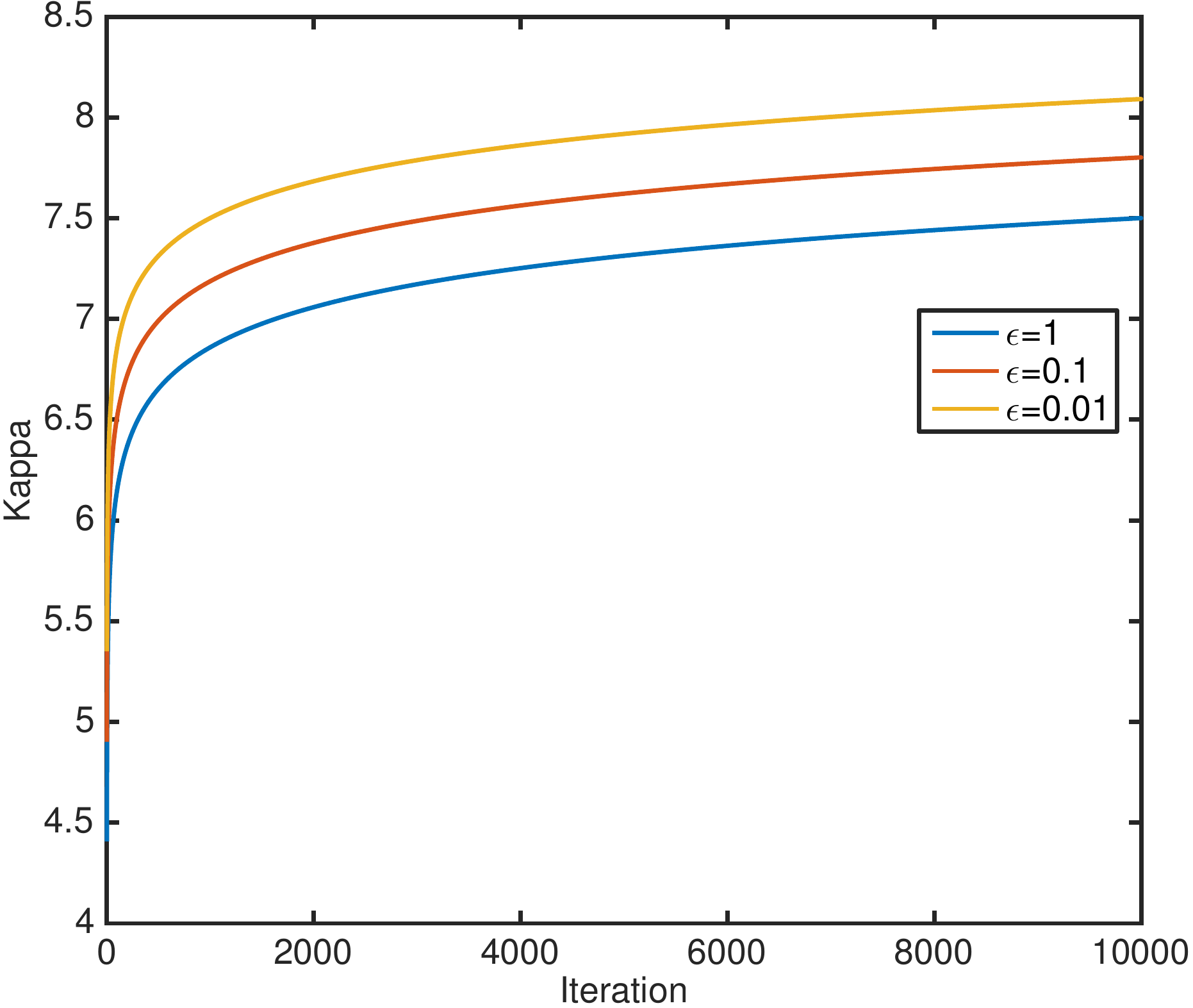}		
		\caption{Change of $\kappa$ value over time: it begins with a small value to exploit the mean estimates and it increases over time in order to explore.}
		\label{fig:kappa}
	\end{center}
\end{figure}

Although several different criteria exist in the literature (see \cite{shahriaritaking}), {\small \sf BO4CO} uses \textit{Lower Confidence Bound} (LCB) \cite{shahriaritaking}. LCB selects the next configuration by trade-off between exploitation and exploration:
\begin{align}  \label{eq:lcb}
u_{LCB}(\mathbf{x}|\mathcal{M}, \mathbb{S}_{1:n})=\argmin_{\mathbf{x}\in \mathbb{X}} \mu_t(\mathbf{x}) - \kappa\sigma_t(\mathbf{x}),
\end{align}
where $\kappa$ can be set according to the objectives. For instance, if we require to find a near optimal configuration quickly we set a low value to $\kappa$ to take the most out of the initial design knowledge. However, if we want to skip local minima, we can set a high value to $\kappa$. Furthermore, $\kappa$ can be adapted over time to benefit from the both \cite{Jamshidi-fql}. For instance, $\kappa$ can start with a reasonably small value to exploit the initial design and increase over time to do more explorations (cf. Figure \ref{fig:kappa}).


%

\subsection{Illustration}

\begin{figure}[t]
	\begin{center}
		\includegraphics[width=8.5cm]{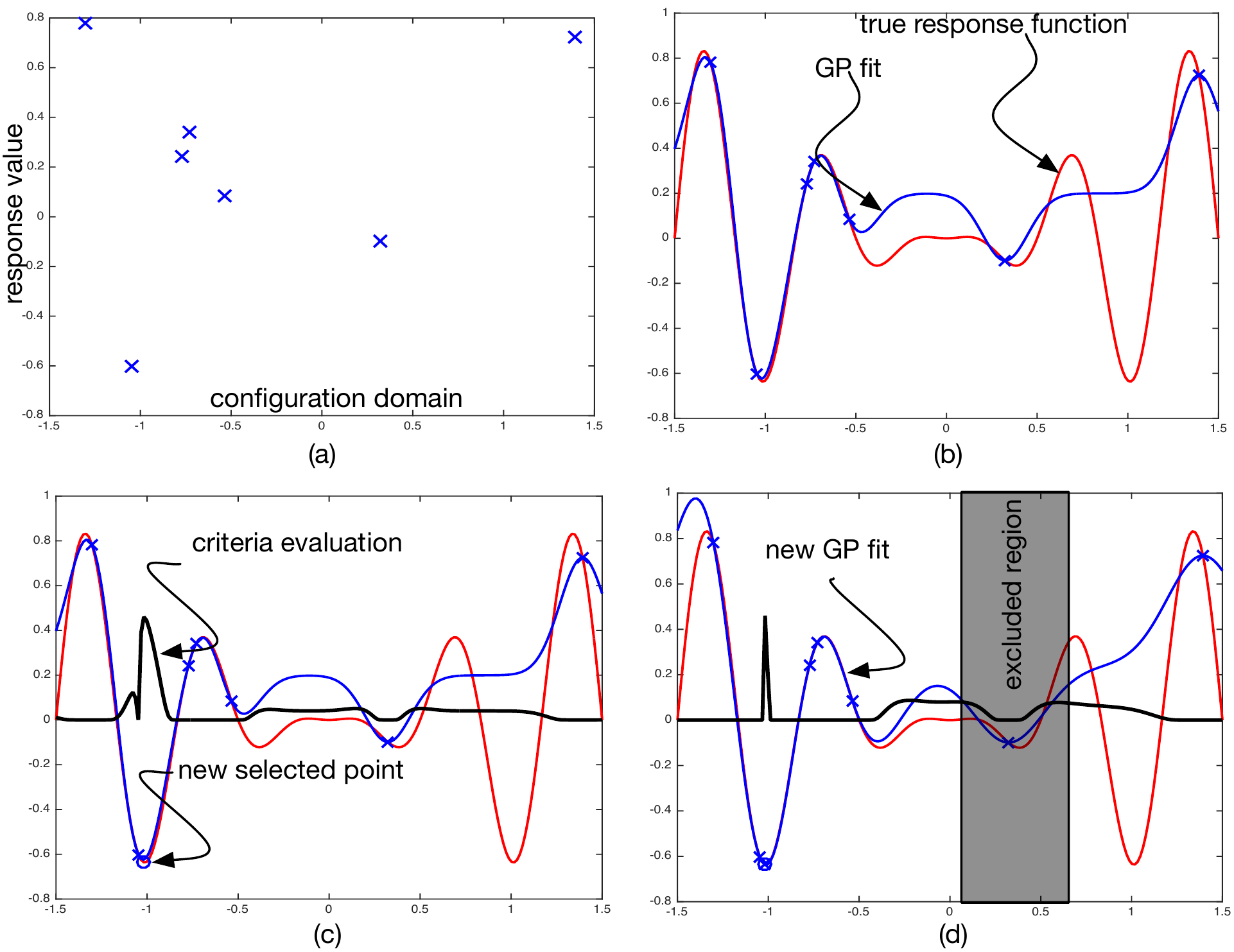}
		\caption{Illustration of configuration parameter optimization: (a) initial observations; (b) a GP model fit; (c) choosing the next point; (d) refitting a new GP model.}
		\label{fig:illustration-gp}
	\end{center}
\end{figure}

The steps in Algorithm~\ref{alg:bo-alg} are illustrated in Figure \ref{fig:illustration-gp}. Firstly, an initial design based on lhd is produced (Figure \ref{fig:illustration-gp}(a)). Secondly, a GP model is fit to the initial design (Figure \ref{fig:illustration-gp}(b)). Then, the model is used to calculate the selection criteria (Figure \ref{fig:illustration-gp}(c)). Finally, the configuration that maximizes the selection criteria is used to run the next experiment and provide data for refitting a more accurate model (Figure \ref{fig:illustration-gp}(d)). 

\subsection{Model fitting in BO4CO}
In this section, we provide some practical considerations to make GPs applicable for configuration optimization.

\subsubsection{Kernel function}
\label{sec:kernel}


In {\small\sf BO4CO}, as shown in Algorithm \ref{alg:bo-alg}, the covariance function $k:\mathbb{X}\times\mathbb{X}\rightarrow\mathbb{R}$ dictates the structure of the response function we fit to the observed data. For integer variables (cf. Section \ref{sec:problem}), we implemented the Mat\'ern kernel \cite{gpml}. The main reason behind this choice is that along each dimension in the configuration response functions different level of smoothness can be observed (cf. Figure \ref{fig:response-latency-wc-motivation}). Mat\'ern kernels incorporate a smoothness parameter $\nu>0$ that permits greater flexibility in modeling such functions \cite{gpml}. 
The following is a variation of the Mat\'ern kernel for $\nu=1/2$:
\begin{align}
\label{eq:matern}
k_{\nu=1/2}(\mathbf{x}_i,\mathbf{x}_j)&= \theta_0^2\exp(-r),
\end{align}
where $r^2(\mathbf{x}_i,\mathbf{x}_j)=(\mathbf{x}_i-\mathbf{x}_j)^\intercal \mathbf{\Lambda} (\mathbf{x}_i-\mathbf{x}_j)$ for some positive semidefinite matrix $\mathbf{\Lambda}$. For categorical variables, we implemented the following \cite{hutter2009automated}:
\begin{align}
\label{eq:categorical}
k_{\theta}(\mathbf{x}_i,\mathbf{x}_j)=\exp(\Sigma_{\ell=1}^{d}(-\theta_\ell\delta(\mathbf{x}_i\neq\mathbf{x}_j))),
\end{align}
where $d$ is the number of dimensions (\emph{i.e.}, the number of configuration parameters), $\theta_\ell$ adjust the scales along the function dimensions and $\delta$ is a function gives the distance between two categorical variables using Kronecker delta \cite{hutter2009automated,shahriaritaking}.
{\sf TL4CO} uses different scales $\{\theta_\ell, \ell=1\dots d\}$ on different dimensions as suggested in \cite{gpml,shahriaritaking}, this technique is called Automatic Relevance Determination (ARD).
After learning the hyper-parameters (\emph{step 6}), if the $\ell$-th dimension turns out to be irrelevant, then $\theta_\ell$ will be a small value, and therefore, will be discarded. This is particularly helpful in high dimensional spaces, where it is difficult to find the optimal configuration. 




\subsubsection{Prior mean function}
\label{sec:prior}

While the kernel controls the structure of the estimated function, the prior mean $\mu(\mathbf{x}):\mathbb{X}\rightarrow\mathbb{R}$ provides a possible offset for our estimation. By default, this function is set to a constant $\mu(\mathbf{x}):=\mu$, which is inferred from the observations \cite{shahriaritaking}. However, the prior mean function is a way of incorporating the expert knowledge, if it is available, then we can use this knowledge. Fortunately, we have collected extensive experimental measurements and based on our datasets (cf. Table \ref{tab:configuration-parameters}), we observed that typically, for Big Data systems, there is a significant distance between the minimum and the maximum of each function (cf. Figure \ref{fig:response-latency-wc-motivation}). Therefore, a linear mean function $\mu(\mathbf{x}):=\mathbf{a}\mathbf{x}+b$, allows for more flexible structures, and provides a better fit for the data than a constant mean. We only need to learn the slope for each dimension and an offset (denoted by $\mu_\ell=(\mathbf{a},b)$).



\subsubsection{Learning parameters: marginal likelihood}
\label{sec:likelihood}

This section describe the \emph{step 7} in Algorithm \ref{alg:bo-alg}. Due to the heavy computation of the learning, this process is computed only every $N_l$ iterations. 
For learning the hyper-parameters of the kernel and also the prior mean functions (cf. Sections \ref{sec:kernel} and \ref{sec:prior}), we maximize the marginal likelihood  \cite{shahriaritaking} of the observations $\mathbb{S}_{1:t}$. To do that, we train GP model \eqref{eq:gp-surrogate-mean-sigma} with $\mathbb{S}_{1:t}$.
We optimize the marginal likelihood using multi-started quasi-Newton hill-climbers \cite{rasmussen2010gaussian}. For this purpose, we use the off-the-shelf {\sf gpml} library presented in \cite{rasmussen2010gaussian}. Using the kernel defined in \eqref{eq:categorical}, we learn $\boldsymbol{\theta}:=(\theta_{0:d},\mu_{0:d},\sigma^2)$ that comprises the hyper-parameters of the kernel and mean functions. The learning is performed iteratively resulting in a sequence of $\boldsymbol{\theta}_i$ for $i=1\dots\lfloor \frac{N_{max}}{N_\ell} \rfloor$. 


\subsubsection{Observation noise}

The primary way for determining the noise variance $\sigma$ in {\small \sf BO4CO} is to use historical data: In Section \ref{sec:uncertain-measurement}, we have shown that such noise can be measured with a high confidence and the signal-to-noise ratios shows that such noise is stationary. The secondary alternative is to learn the noise variance sequentially as we collect new data. We treat them just as any other hyper-parameters, see Section \ref{sec:likelihood}.



\begin{figure}[t]
	\begin{center}
		\includegraphics[width=5.5cm]{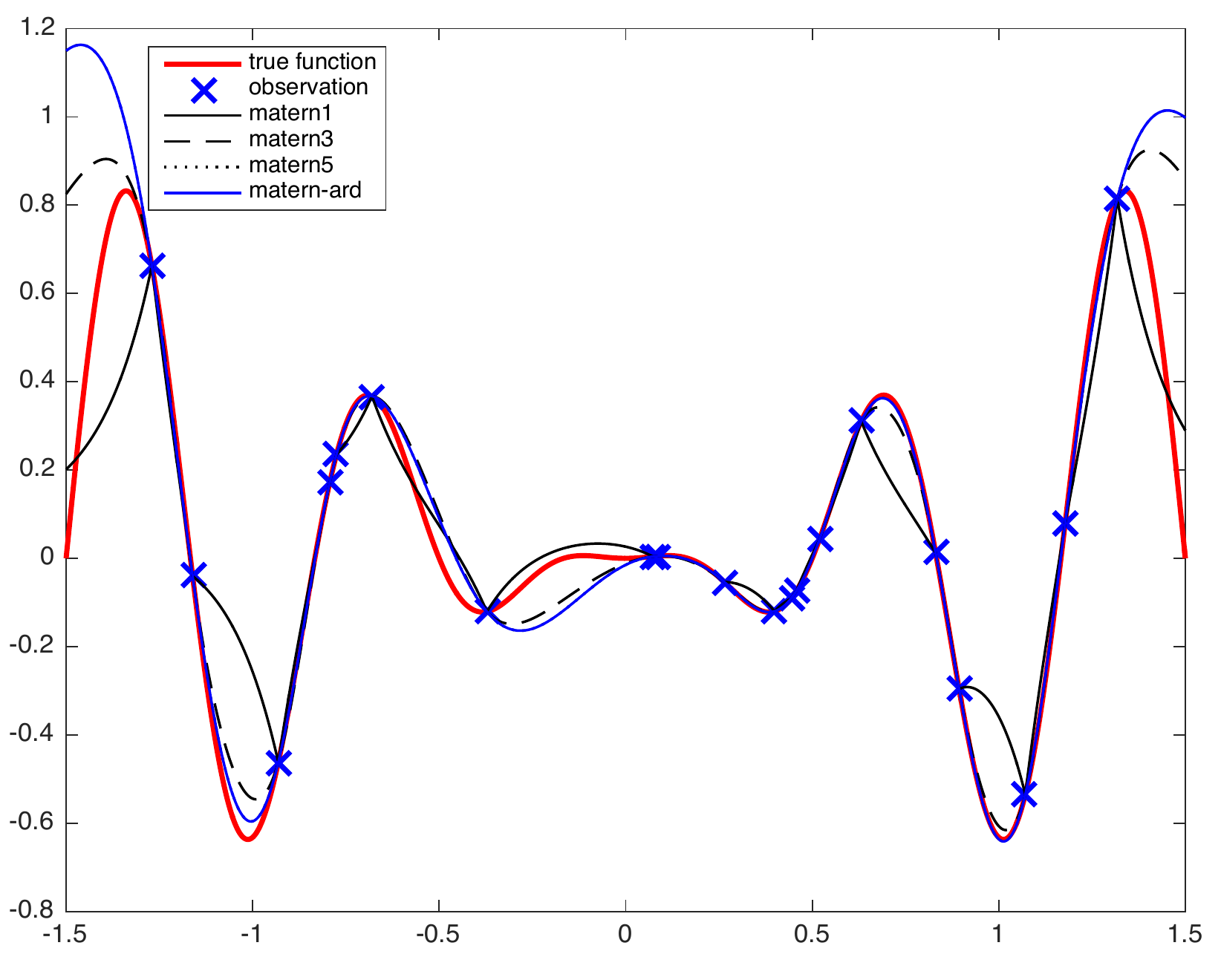}
		\caption{The effect of changing the kernel in estimations.}
		\label{fig:kernel-profiles}
	\end{center}
\end{figure}

%% file: sections/experiment.tex
\section{Experimental Results}
\label{sec:experiments}

\subsection{Implementation}
\label{sec:implementation}
From an implementation perspective, {\small \sf BO4CO} consists of three major components: (i) an \emph{optimization component} (cf. left part of Figure \ref{fig:configuration-optimizer}), (ii) an \emph{experimental suite} (cf. right part of Figure \ref{fig:configuration-optimizer}) integrated via a (iii) \emph{data broker}. The optimization component implements the model (re-)fitting \eqref{eq:gp-surrogate-mean-sigma} and criteria optimization \eqref{eq:next-point} steps in Algorithm \ref{alg:bo-alg} and is developed in Matlab 2015b. The experimental suite component implements the facilities for automated deployment of topologies, performance measurements and data preparation and is developed in Java. The optimization component retrieves the initial design performance data and determines which configuration to try next using the procedure explained in \ref{sec:acquision-functions}. The suite then deploys the \emph{topology under test} on a testing cluster. The performance of the topology is then measured and the performance data will be used for model refitting. We have released the \emph{code} and \emph{data}: \url{https://github.com/dice-project/DICE-Configuration-BO4CO}.

In order to make  {\small \sf BO4CO} more practical and relevant for industrial use, we considered several implementation enhancements. In order to perform efficient GP model re-fitting, we implemented a covariance wrapper function that keeps the internal state for caching kernels and its derivatives, and can update kernel function by a single element. This was particularly helpful for learning the hyper-parameters at runtime. 

\subsection{Experimental design}
\label{sec:experiment-design}

\subsubsection{Topologies under test and benchmark functions}
In this section, we evaluate {\small \sf BO4CO} using 3 different Storm benchmarks: (i) {\sf WordCount}, (ii) {\sf RollingSort}, (iii) {\sf SOL}. {\sf RollingSort} implements a common pattern in real-time data analysis that performs rolling counts of incoming messages. {\sf RollingSort} is used by Twitter for identifying trending topics. 
{\sf SOL} is a \emph{network intensive} topology, where the incoming messages will be routed through an inter-worker network. 
{\sf WordCount} and {\sf RollingSort} are standard benchmarks and are widely used in the community, \emph{e.g.}, research papers \cite{ghazal2013bigbench} and industry scale benchmarks \cite{huang2010hibench}.
We have conducted all the experiments on 5 cloud clusters and with different sets of parameters resulted in datasets in Table \ref{tab:configuration-parameters}.

We also evaluate {\small \sf BO4CO} with a number of benchmark functions, where we perform a synthetic experiment inside MATLAB in which a measurement is just a function evaluation: {\sf Branin(2D), Dixon-Szego(2D), Hartmann(3D)} and {\sf Rosenbrock(5D)}. 
These benchmark functions are commonly used in global optimization and configuration approaches \cite{xi2004smart, thonangi2008finding}. 
We particularly selected these because: (i) they have different curvature and (ii) they have multiple global minimizers, and (iii) they are of different dimensions. 


%
%


\subsubsection{Baseline approaches}
\label{sec:baselines}
The performance of {\small \sf BO4CO} is compared with the 5 outstanding state-of-the-art approaches for configuration optimizations: {\sf SA} \cite{guo2010evaluating}, {\sf GA} \cite{behzad2013taming}, {\sf HILL} \cite{xi2004smart}, {\sf PS} \cite{thonangi2008finding} and {\sf Drift} \cite{sun2013random}. They are of different nature and use different search algorithms: simulated annealing, genetic algorithm, hill climbing, pattern search and adaptive search. 



\subsubsection{Experimental considerations}
\label{sec:experiment-considerations}
The performance statistics regarding each specific configuration has been collected over a window of 5 minutes (excluding the first two minutes of burn-in and the last minute of cluster cleaning). The first two minutes are excluded because the monitoring data are not stationary, while the last minute is the time given to the topology to fully process all messages. We then shut down the topology, clean the cluster and move on to the next experiment. We also replicated each runs of algorithms for 30 times in order to report the comparison results. Therefore, all the results presented in this paper are the mean performance over 30 runs. 

\subsubsection{Cluster configuration}
\label{sec:cluster-configuration}
We conducted all the experiments on 5 different multi-node clusters on three various cloud platforms, see \ref{sec:appendix} for more details. The reason behind this decision was twofold: (i) saving time in collecting experimental data by running topologies in parallel as some of the experiments supposed to run for several weeks, see Section \ref{sec:parameters-interactions}. (ii) replicating the experiment with different processing node. 


%
%
%

%

\subsection{Experimental analysis}
\label{sec:experimental-analysis}
In the following, we evaluate the performance of each approach as a function of the number of evaluations. So for each case, we report performance using the absolute distance of the \emph{minimum function value} from the global minimum. Since we have measured all combinations of parameters in our datasets, we can measure this distance at each iteration.

\subsubsection{Benchmark functions global optimization}
\label{sec:benchmark-functions}
\begin{figure}[t] 
 	\begin{center}
 		\includegraphics[width=\columnwidth]{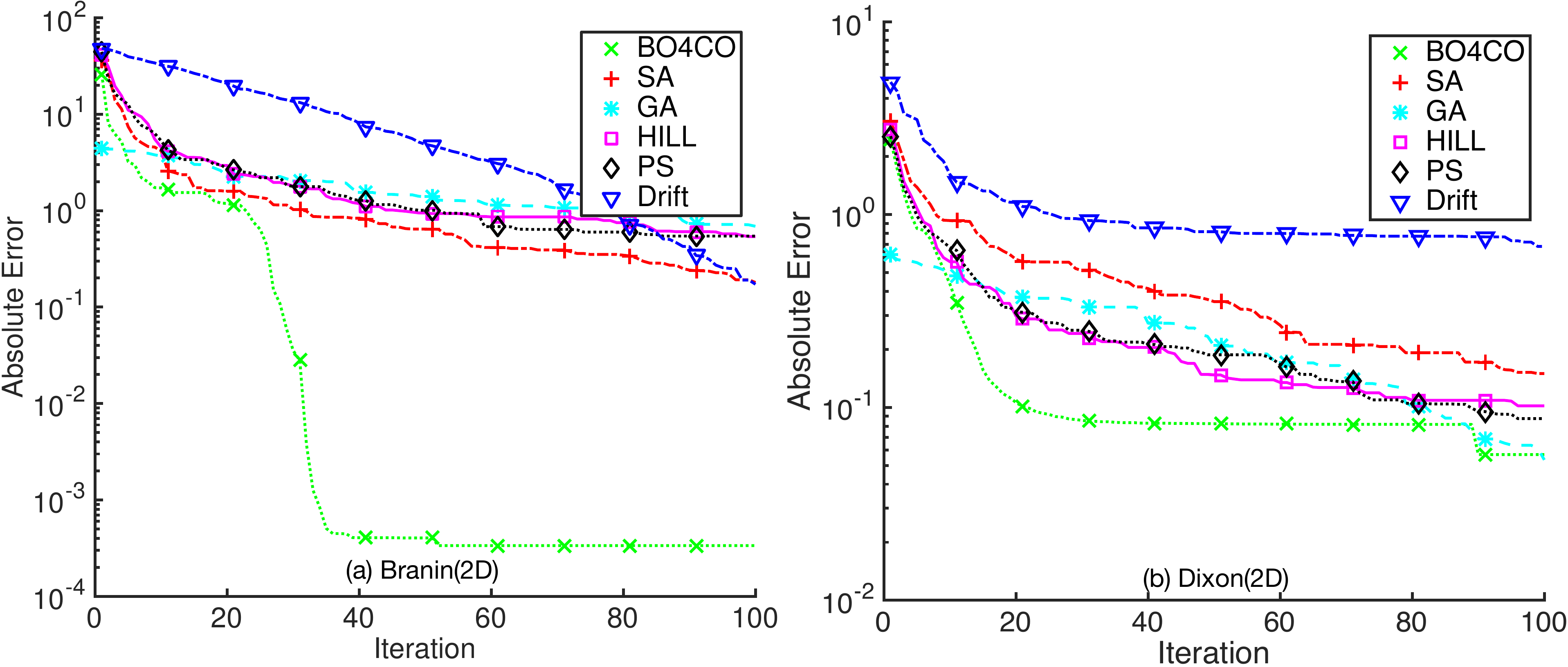}
 		\caption{{\sf Branin(2D), Dixon(2D)} test function optimization.}
 		\label{fig:absolute-error-branin-dixon}
 	\end{center}
 \end{figure}

\begin{figure}
 	\begin{center}
 		\includegraphics[width=6.5cm]{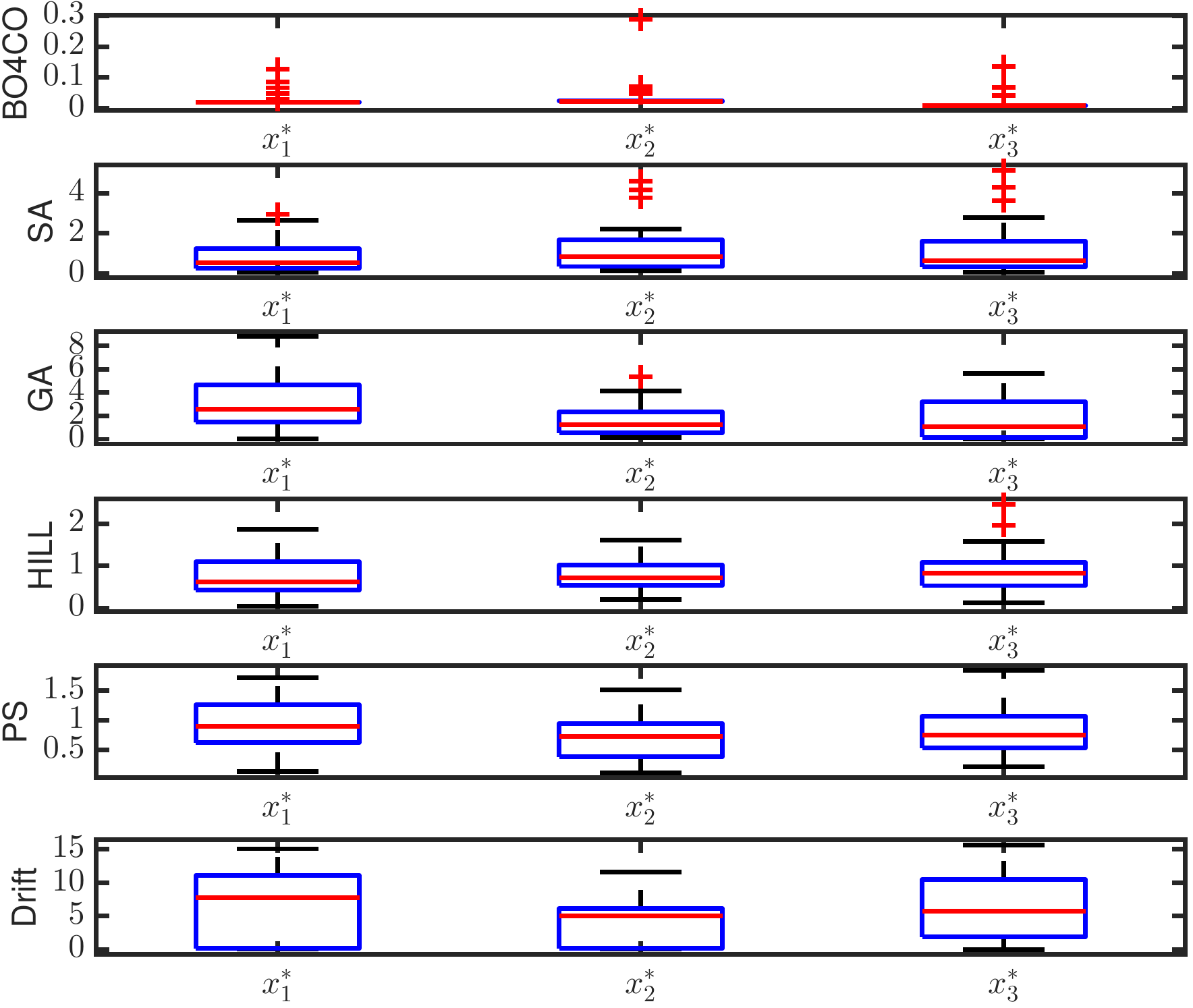}
 		\caption{Distance to the three {\sf Branin}'s minimizers.}
 		\label{fig:distance2minimizers-branin}
 	\end{center}
 \end{figure}

\begin{figure}[t] 
 	\begin{center}
 		\includegraphics[width=\columnwidth]{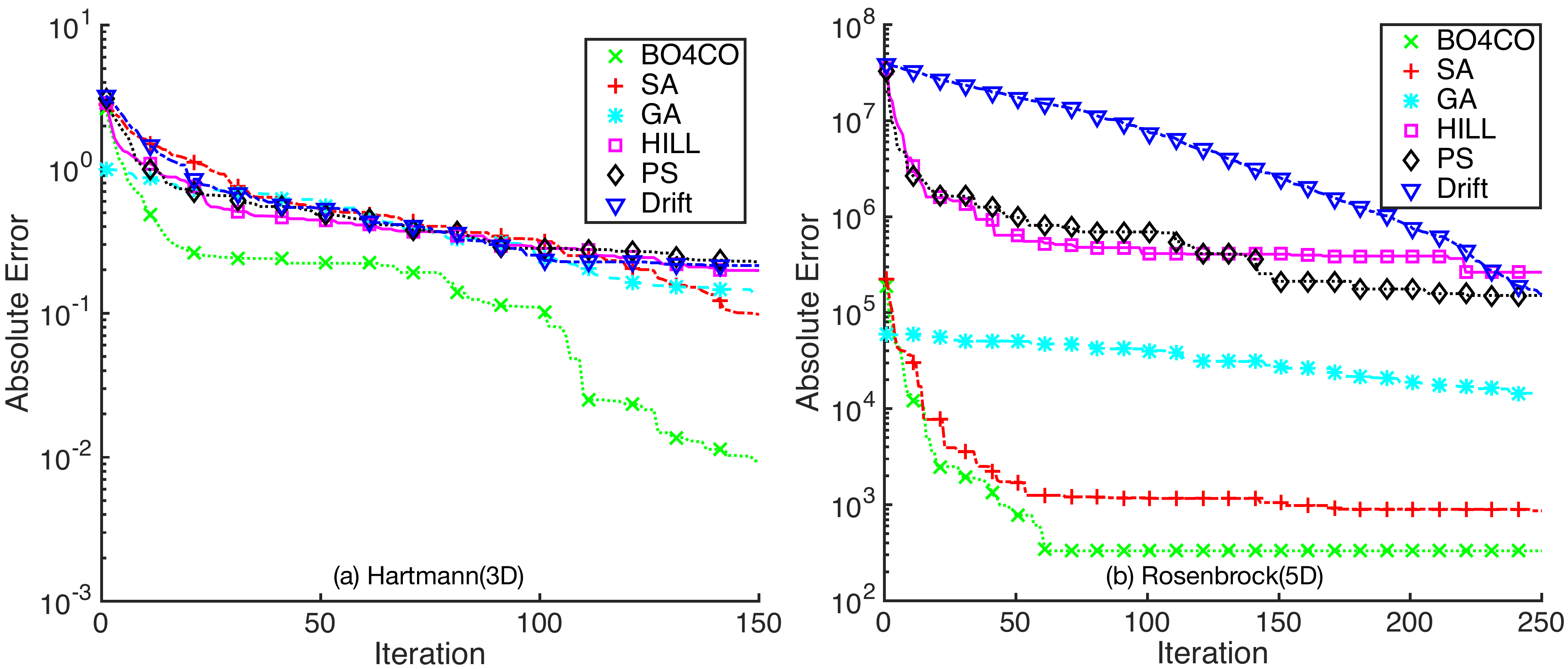}
 		\caption{{\sf Hartmann(3D), Rosenbrock(5D)} optimization.}
 		\label{fig:absolute-error-hartmann3-rosen5}
 	\end{center}
 \end{figure}

The results for {\sf Branin} in Figure \ref{fig:absolute-error-branin-dixon}(a) show that {\small \sf BO4CO} outperforms the other approaches with three orders of magnitude, while this gap is only an order of magnitude for {\sf Dixon} as in Figure \ref{fig:absolute-error-branin-dixon}(b). This difference can be associated to the fact that {\sf Dixon} surface is more rugged than {\sf Branin} \cite{weise2009global}. For {\sf Branin}, {\small \sf BO4CO} finds the global minimum within the first 40 iterations, while even {\sf SA} stalls on a local minimum. The rest, including {\sf GA}, {\sf HILL}, {\sf PS} perform similarly to each other throughout the experiment and reach a local minimum, which {\small \sf BO4CO} finds only within the first 10 iterations. For {\sf Dixon}, {\small \sf BO4CO} gets close to the global minimum within the first 20 iterations, while the best performers (\emph{i.e.}, {\sf PS} and {\sf HILL}) approach to points an order of magnitude away comparing with {\small \sf BO4CO}. 
{\sf Branin} function has 3 global minimizers at $x_1^* = (−\pi, 12.27), x_2^* = (\pi, 2.27), x_3^* = (9.42,2.47)$, interestingly comparing to baselines, {\small \sf BO4CO} gets close to all minimizers (cf. Figure \ref{fig:distance2minimizers-branin}). Therefore, {\small \sf BO4CO} gains information on all minimizers as opposed to baselines. 
The results for {\sf Hartmann} in Figure \ref{fig:absolute-error-hartmann3-rosen5}(a) show that {\small \sf BO4CO} decreases the absolute error quickly after 20 iterations, but only approaches to the global minimum after 120 iterations. Neither of the baseline approaches get close to the global minimum even after 150 iterations. 
The good performance of {\small \sf BO4CO} is also confirmed in the case of {\sf Rosenbrock}, as shown in Figure \ref{fig:absolute-error-hartmann3-rosen5}(b). {\small \sf BO4CO} finds the optimum in such large space only after 60 iterations, while {\sf GA}, {\sf HILL}, {\sf PS} and {\sf Drift} perform poorly with an error of three orders of magnitude higher than our approach. However, {\sf SA} performs well with an order of magnitude away from the ones found by {\small \sf BO4CO}. 


%
%
%
 
 
\subsubsection{Storm configuration optimization}
\label{sec:storm-dataset-experiment}
We now discuss the results of {\small \sf BO4CO} on the Storm datasets in Table \ref{tab:configuration-parameters}: {\sf SOL(6D)}, {\sf RollingSort(6D)},  {\sf WordCount(3D,5D)}.

\begin{figure}[t]
	\begin{center}
		\includegraphics[width=\columnwidth]{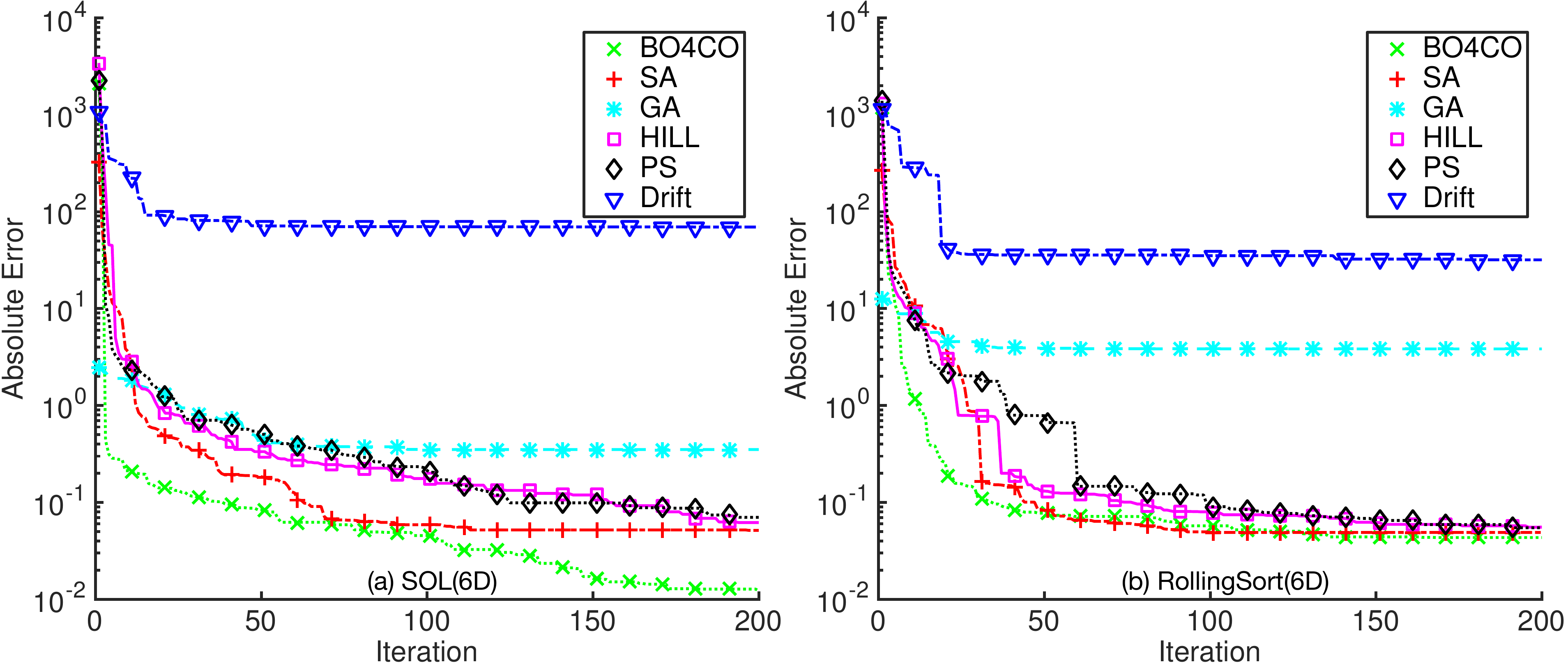}		\caption{{\sf SOL(6D), RollingSort(6D)} optimization.}
		\label{fig:c3-c4}
	\end{center}
\end{figure}

The results for {\sf SOL} in Figure \ref{fig:c3-c4}(a) show that {\small \sf BO4CO} decreases the optimality gap within the first 10 iterations and decreases this gap until iteration 200 and does not get trapped into a local minimum. Instead, baseline approaches like {\sf Drift} and {\sf GA} get trapped into a local minimum in early iterations, while {\sf HILL} and {\sf PS} get stuck some iterations later at 120. Among the baselines, {\sf SA} performs the best and it decreases the optimality gap in the first 70 iterations, however, it gets stuck to a local optimum thereafter. 

The results for {\sf RollingSort} in Figure \ref{fig:c3-c4}(b) are similar to the {\sf SOL} ones. {\small \sf BO4CO} decreases the error considerably in the first 50 iterations, while the baseline approaches, except {\sf SA} and {\sf HILL}, perform poorly in that period. However, during iterations 100-200, the rests, except {\sf GA} and {\sf Drift}, find configuration with close performance as the ones {\small \sf BO4CO} finds.

\begin{figure}[t]
	\begin{center}
		\includegraphics[width=\columnwidth]{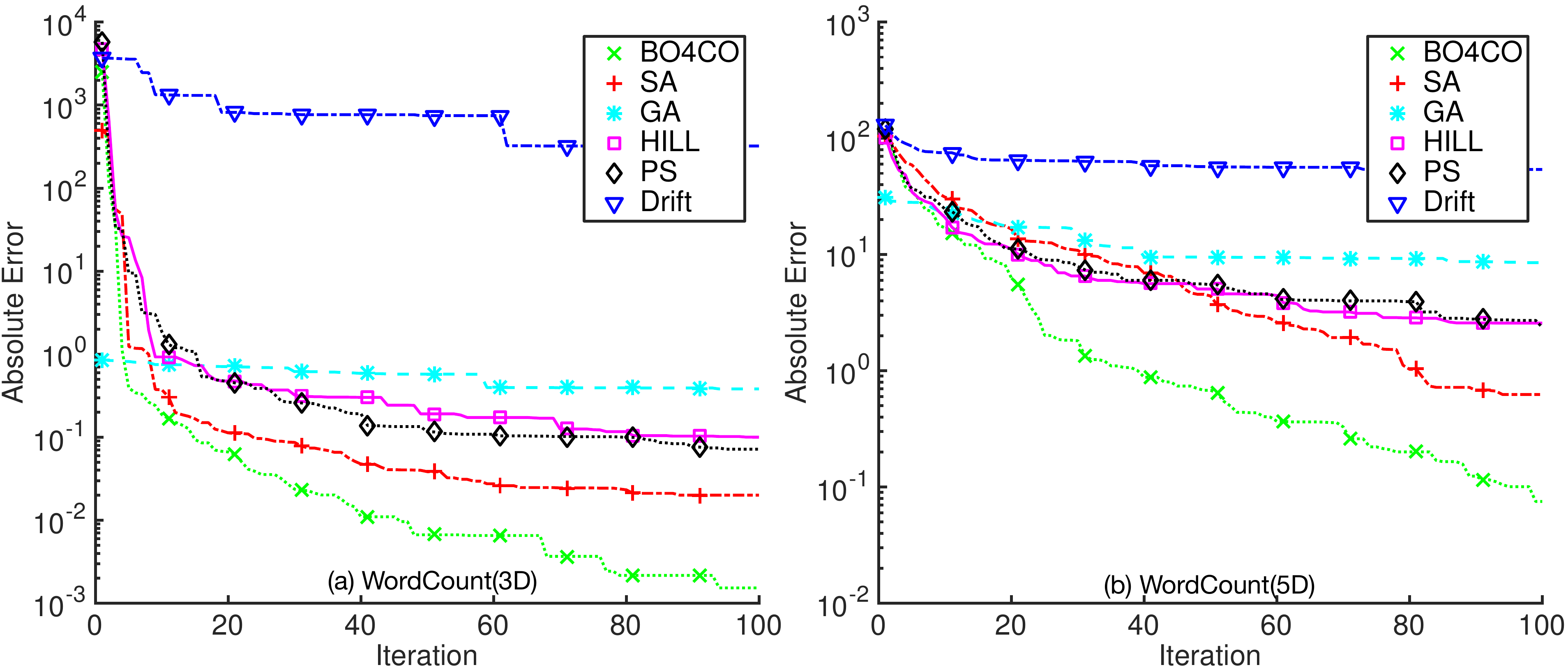}		
		\caption{{\sf WordCount(3D,5D)} configuration optimization.}
		\label{fig:c1-c5}
	\end{center}
\end{figure}

For {\sf WordCount} (Figure \ref{fig:c1-c5}(a,b)) the results are different. {\small \sf BO4CO} outperforms the best baseline performer, \emph{i.e.}, {\sf SA}, by an order of magnitude, while the others by at least two orders of magnitude. Among the baselines, {\sf SA} performs the best for {\sf WordCount}(3D), while for {\sf WordCount}(5D) dataset, {\sf HILL} and {\sf PS} performs better in the first 50 iterations.  

Summarizing, while the results for the Storm benchmarks are consistent with the ones we observed for the benchmark functions, it shows a clear gain in favor of {\small \sf BO4CO}, with at least an order of magnitude in the initial iterations. In each case, {\small \sf BO4CO} finds a better configuration much more quickly than baselines. As opposed to the benchmark functions, {\sf SA} consistently outperforms the rest of baseline approaches. To highlight this achievement, note that 50 iterations for a dataset like {\sf SOL (6D)} is only 1\% of the total number of possible tests for finding the optimum configurations and identifying such configurations with a latency close to the global optimum can save a considerable time and cost.  


%

%

\subsection{Sensitivity analysis}
\label{sec:sensitivity}

\subsubsection{Prediction accuracy of the learned GP model} 
 \label{sec:acuracy}
\begin{figure}[t]
	\begin{center}
		\includegraphics[width=5cm]{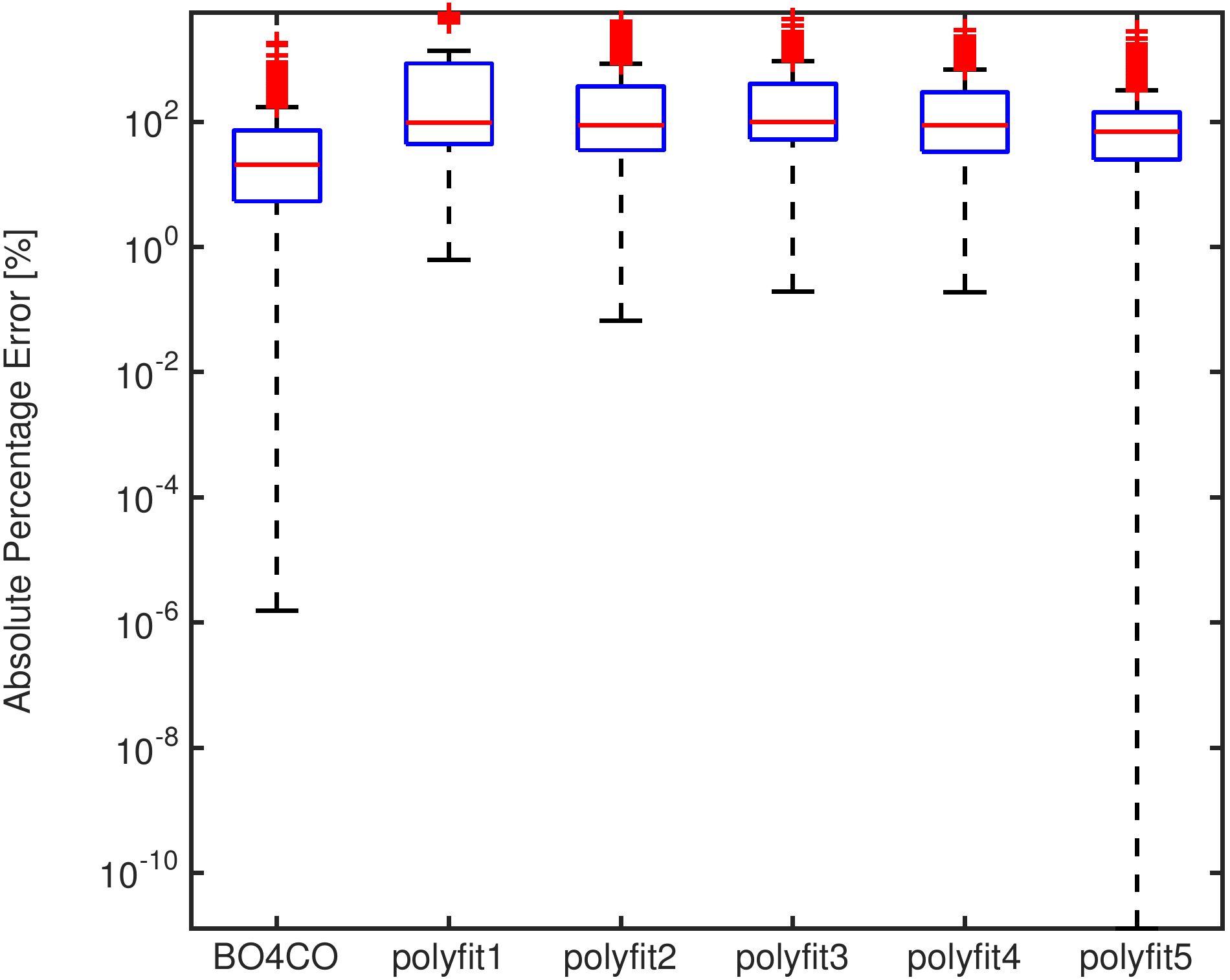}		
		\caption{Absolute percentage error of predictions made by {\small \sf BO4CO}'s GP fit after 100 iterations vs multivariate polynomial regression models for {\sf WordCount(3D)} dataset.}
		\label{fig:model-prediction-comparison}
	\end{center}
\end{figure}

\begin{figure}[t]
 	\begin{center}
 		\includegraphics[width=\columnwidth]{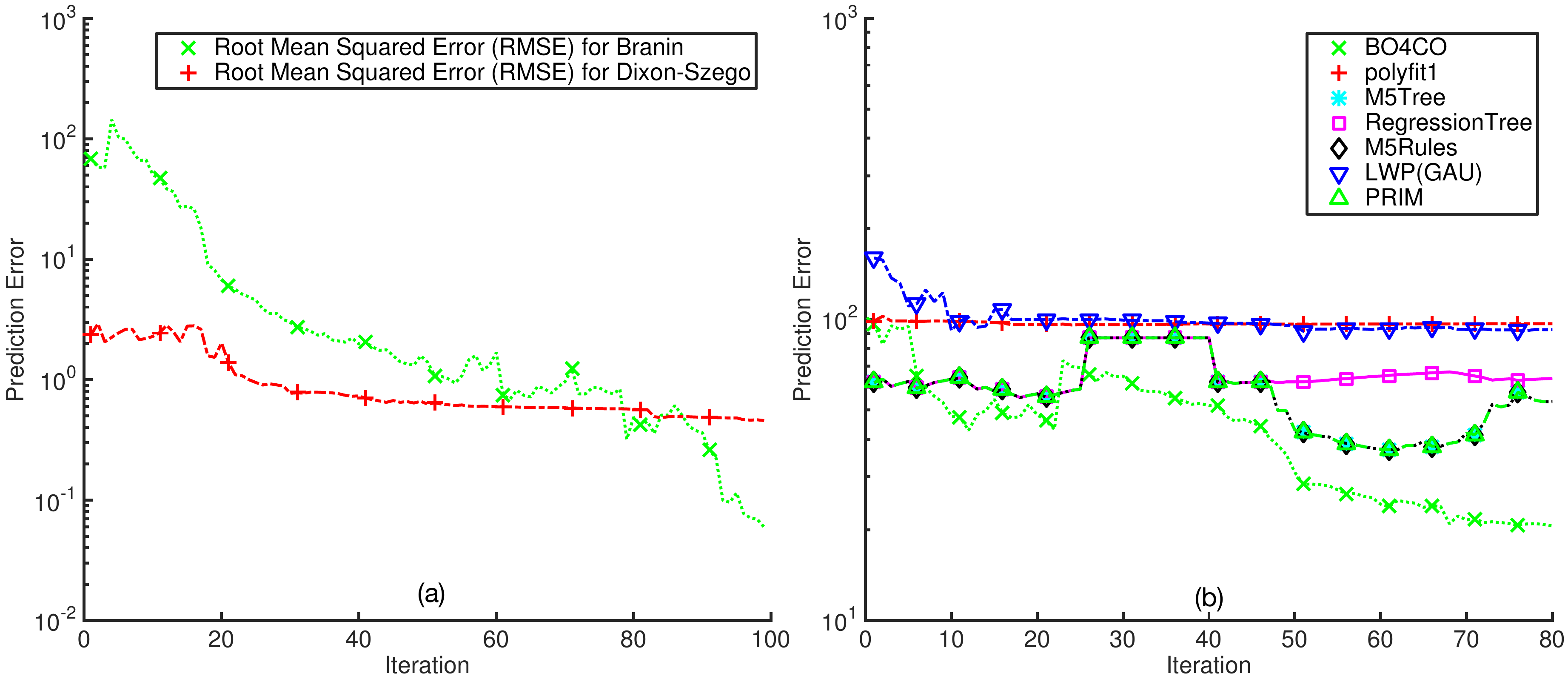}	
 		\caption{(a) improving accuracy of GP models over time, (b) comparing prediction accuracy of GPs with other machine learning models on {\sf wc(6D)}.}
 		\label{fig:branin-dixon-c1-model-prediction-evolution}
 	\end{center}
 \end{figure}

Since {\small \sf BO4CO} does not uniformly sample the configuration space (cf. Figure \ref{fig:example-gp},\ref{fig:illustration-gp}), we speculated that the GP models trained in {\small \sf BO4CO} are not useful for predicting the performance of configurations that have not been experimented. However, when we compared the GP model on the {\sf WordCount}, it was much more accurate than the ones of polynomial regressions (see Figure \ref{fig:model-prediction-comparison}). This shows a clear advantage over design of experiments (DoE), which normally uses first-order and second-order polynomials. The root mean squared error (RMSE) for {\sf Branin} and {\sf Dixon} in \ref{fig:branin-dixon-c1-model-prediction-evolution}(a) clearly show that the GP models can provide accurate predictions after 20 iterations. This fast learning rate can be associated to the power of GPs for regressions \cite{gpml}. We further compared the prediction accuracy of the GP models trained in {\small \sf BO4CO} with several machine learning models (including {\sf M5Tree, Regression Tree, LWP, PRIM} \cite{gpml}) in Figure \ref{fig:branin-dixon-c1-model-prediction-evolution}(b) and we observed that the GP model predictions were more accurate, while the accuracy of other models either did not improve (\emph{e.g.}, {\sf M5Tree}) or was deteriorated (\emph{e.g.,} {\sf PRIM, polyfit5}) due to over-fitting.

%



\subsubsection{Exploitation vs. exploration}
\label{sec:exploitation-vs-exploration}
\begin{figure}[t]
	\begin{center}
		\includegraphics[width=\columnwidth]{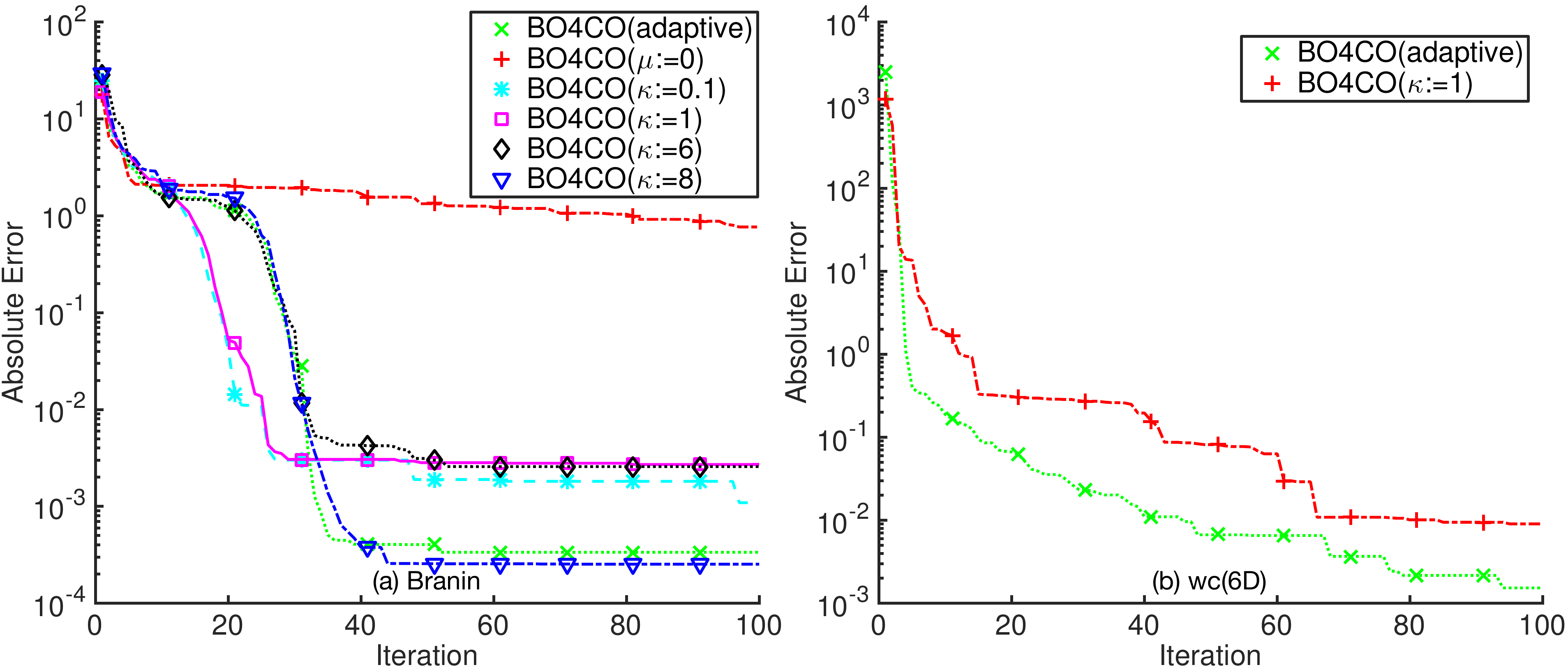}		
		\caption{Exploitation vs exploration (a) {\sf Branin}, (b) {\sf wc(6D)}.}
		\label{fig:exploration-branin-c1}
	\end{center}
\end{figure}

\begin{figure}[t]
	\begin{center}
		\includegraphics[width=5cm]{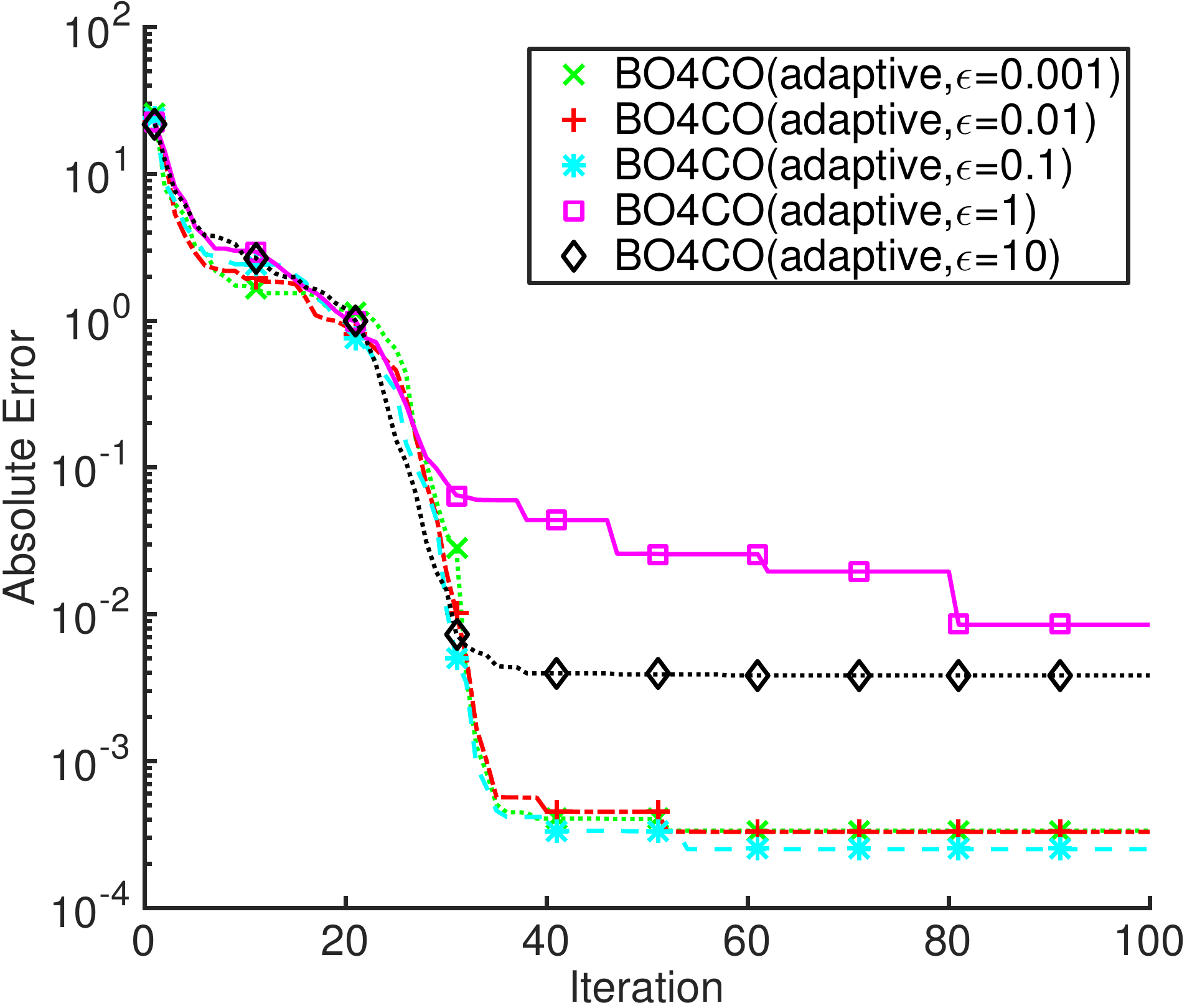}		\caption{Exploitation vs. exploration: compare the rate of changes in $\kappa$ with different value of $\epsilon$ in Figure \ref{fig:kappa}.}
		\label{fig:exploration-epsilon-sensitivity}
	\end{center}
\end{figure}

In \eqref{eq:lcb}, $\kappa$ adjusts the exploitation-exploration: small $\kappa$ means high exploitation, while a large $\kappa$ means a high exploration. The results in Figure \ref{fig:exploration-branin-c1}(a) show that using a relatively high exploration (\emph{i.e.}, $\kappa=8$) performs better up to an order of magnitude comparing with high exploitations (\emph{i.e.}, $\kappa=0.1$). However, for $\kappa=0.1,1$ exploiting the mean estimates improves the performance at early iterations comparing with higher explorations cases as in $\kappa=6$. This observation motivated us to tune $\kappa$ dynamically by using a lower value at early iterations to exploit the knowledge gained through the initial design and set a higher value later on, see Section \ref{sec:acquision-functions}. The result for {\sf WordCount} in Figure \ref{fig:exploration-branin-c1}(b) confirms that adaptive $\kappa$ improves the performance considerably over constant $\kappa$. Figure \ref{fig:exploration-epsilon-sensitivity} shows that when we increase $\kappa$ with a higher rate (cf. Figure \ref{fig:kappa}), it will improve the performance. However, the results in Figure \ref{fig:exploration-branin-c1}(a) suggest using a high value of exploration, this should not be set to an extreme where this makes the mean estimates ineffective. As in Figure \ref{fig:exploration-branin-c1}(a), it performs 4 orders of magnitude worse when we ignore the mean estimates (\emph{i.e.}, $\mu_t:=0$).

%
%

\subsubsection{Bootstrapping vs no bootstrapping}  
\label{sec:bootstrapping}
\begin{figure}[t]
	\begin{center}
		\includegraphics[width=\columnwidth]{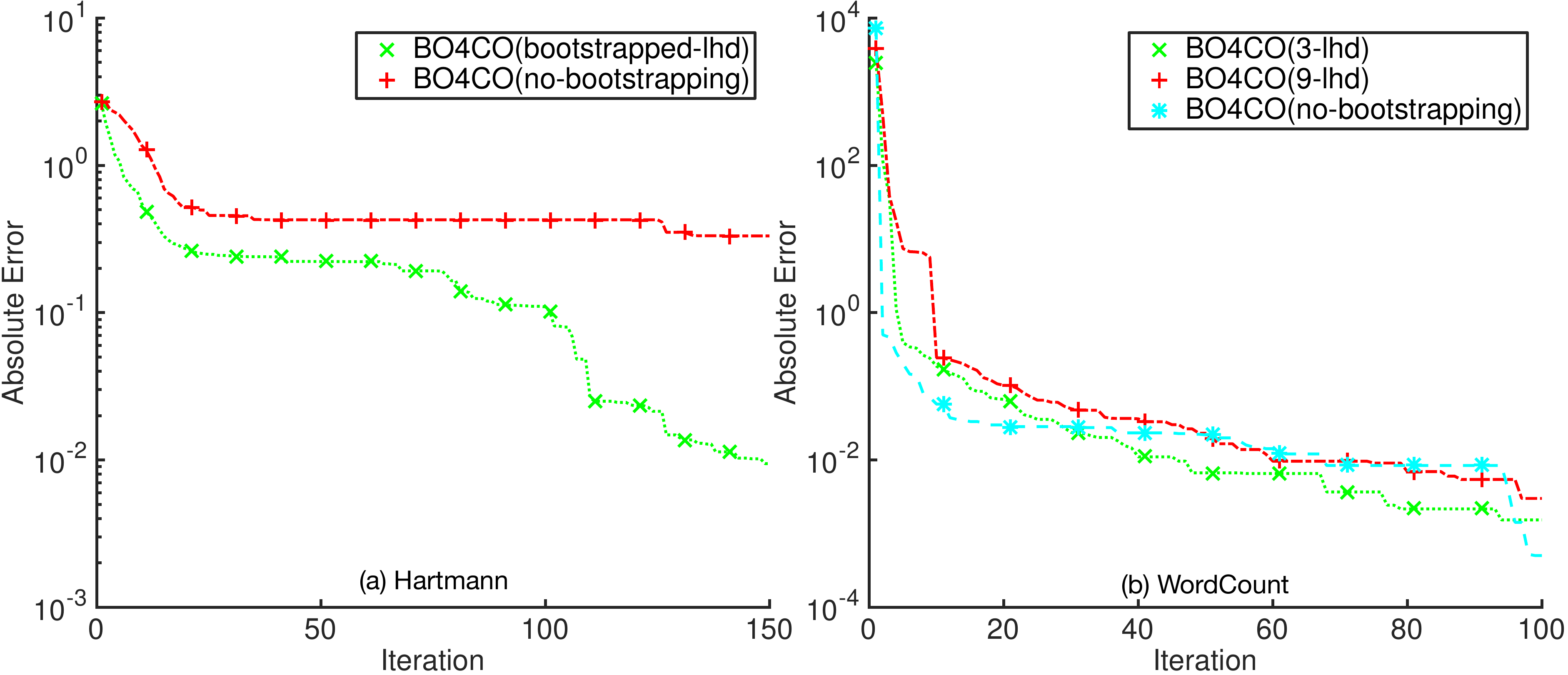}		
		\caption{Bootstrapping vs no bootstrapping: acceleration of performance due to bootstrapping.}
		\label{fig:hartmann3-wc-bootstrap-vs-nobootstrap}
	\end{center}
\end{figure}

{\small \sf BO4CO} uses lhd design in order to bootstrap the search process. The results for {\sf Hartmann} and {\sf WordCount} in Figure \ref{fig:hartmann3-wc-bootstrap-vs-nobootstrap}(a,b) confirms that this choice provides a good opportunity in order to explore along all dimensions and not to trap into local optimum thereafter. However, the results in Figure \ref{fig:hartmann3-wc-bootstrap-vs-nobootstrap}(b) suggest that a high number of initial design may deteriorate the goal of finding the optimum early. 

%% file: sections/discussion.tex
\section{Discussions}
\label{sec:discussion}

\subsection{Computational and memory requirements}
\label{sec:complexity}
\begin{figure}[t]
	\begin{center}
		\includegraphics[width=5cm]{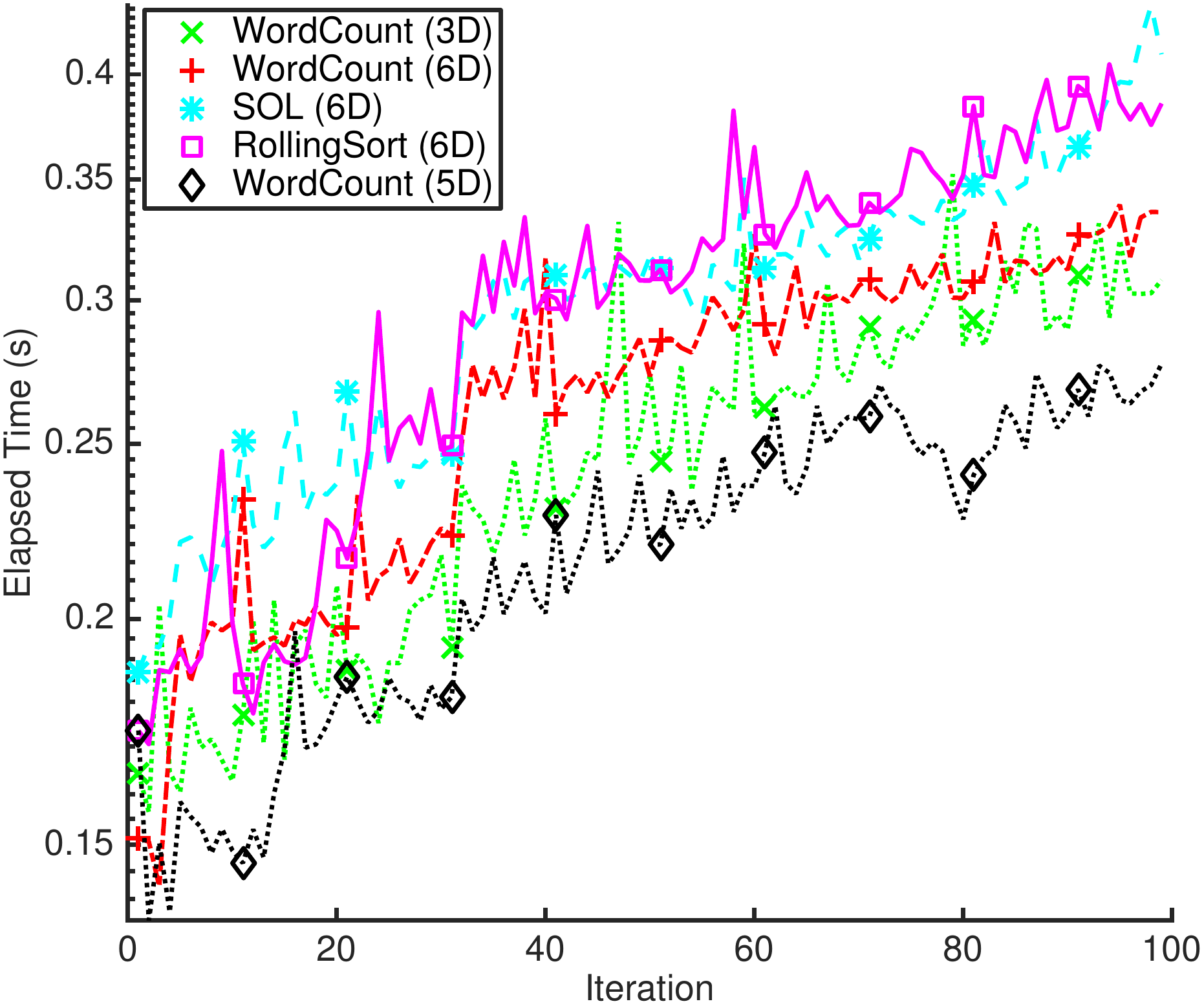}		
		\caption{Runtime overhead of {\small \sf BO4CO} (excluding the experiment time) is in the scale of few hundred milliseconds.}
		\vspace{-1em}
		\label{fig:runtime}
	\end{center}
\end{figure}

The exact inference {\small \sf BO4CO} uses for fitting a GP model to the $t$ observed data is $O(t^3)$ because of inversion of kernel $\mathbf{K}^{-1}$ in \eqref{eq:gp-surrogate-mean-sigma}. We could in principle compute the Cholesky decomposition and use it for subsequent predictions, which would lower the complexity to $O(t^2)$. However, since in {\small \sf BO4CO} we learn the kernel hyper-parameters every $N_\ell$ iterations, Cholesky decomposition must be re-computed, therefore the complexity is in principle $O(t^2\times t/N_\ell)$, where the additional factor of $t/N_\ell$ counts the expected number of iterations. Figure \ref{fig:runtime} provides the computation time for finding the next configuration in Algorithm \ref{alg:bo-alg} for 5 datasets in Table \ref{tab:configuration-parameters}. The time is measured running {\small \sf BO4CO} on a MacBook Pro with 2.5 GHz Intel Core i7 CPU and 16GB of Memory. The computation time in larger datasets ({\sf RollingSort(6D), SOL(6D), WordCount(6D)}) is higher than those with less data and lower dimensions ({\sf WordCount(3,5D)}). Moreover, the computation time increases over time since the matrix size for Cholesky inversion gets larger. 

{\small \sf BO4CO} requires to store 3 vectors of size $|\mathbb{X}|$ for mean, variance and LCB estimates and a matrix of size $|\mathbb{S}_{1:t}|\times |\mathbb{S}_{1:t}|$ for $\mathbf{K}$ and of size $|\mathbb{S}_{1:t}|$ for observations, making the memory requirement of $O(3|\mathbb{X}|+|\mathbb{S}_{1:N_{max}}|( |\mathbb{S}_{1:N_{max}}|+1))$ in total.


\subsection{BO4CO in practice}
\label{sec:behind-the-scene}

{ \em Extensibility.} We have integrated {\sf BO4CO} with continuous integration, delivery, deployment and monitoring tools in a DevOps pipeline as a part of H2020 DICE project. {\sf BO4CO} performs as the configuration tuning tool for Big Data systems. 

{\em Usability.} {\sf BO4CO} is easy to use, end users only need to determine the parameters of interests as well as experimental parameters and then the tool automatically sets the optimized parameters. Currently, {\sf BO4CO} supports Apache Storm and Cassandra. However, it is designed to be extensible. 

{\em Scalability.} The scalability bottleneck is experimentation. The running time of the cubic algorithm is of the order of milliseconds (cf. Figure \ref{fig:runtime}). Each of these experiments takes more than 10 minutes, orders of magnitude over {\sf BO4CO}. 

%% file: sections/relatedwork.tex
\section{Related Work}
\label{sec:related-work}

There exist several categories of approaches to address the system configuration problem, as listed in Table~\ref{tab:category-of-approaches}. 

\emph{Rule-based}: In this category, domain experts create a repository of rules that is able to recommend a good configuration. \emph{e.g.}, IBM DB2 Configuration Advisor \cite{db2advisor}. The Advisor asks administrators a series of questions, \emph{e.g.}, does the workload is CPU or memory intensive? Based on the answers, it recommends a configuration. However, for multi-dimensional spaces such as SPS in which the configuration parameters have unknown non-linear relationship, this approach is naive \cite{thonangi2008finding}.

\emph{Design of experiments}: DoE conducts exhaustive experiments for different combinations of parameters in order to find influential factors \cite{Hall1999}. Although DoE is regarded as a classical approach for application configuration, in multi-dimensional spaces performing naive experiments without any sequential feedback from real environment is infeasible and costly.

\emph{Model-based}: This category conducts a series of experiments where each runs the system using a chosen configuration to observe its performance. Each experiment produces a $(x, f(x))$ sample. A (statistical) model can then be trained from these samples and used to find good configurations. However, an exhaustive set of experiments, usually above the limited budget, need to be conducted to provide a representative data sets, otherwise the prediction based on the trained model will not be reliable (cf. Figure \ref{fig:model-prediction-comparison}). White box \cite{menasce2001preserving} and black box models \cite{johnstonperformance,yigitbasi2013towards,Siegmund} have been proposed.

\emph{Search-based}: In this approach, also known as \emph{sequential design}, experiments can be performed sequentially where the next set of experiments is determined based on an analysis of the previous data. In each iteration a (statistical) model is fitted to the data and this model guides the selection of the next configuration. Evolutionary search algorithms such as simulated annealing, recursive random search \cite{ye2003recursive}, genetic algorithm \cite{behzad2013taming}, hill climbing \cite{xi2004smart}, sampling \cite{Siegmund} and Covariance Matrix Adaptation \cite{Saboori2008} have been adopted.

\emph{Learning-based}: There exists some approaches that employ offline and online learning (\emph{e.g.}, reinforcement learning) to enable online system configuration adaptation \cite{Bu2009}. The approaches in this category, as opposed to the other approaches, try to find optimum configurations and adapt it when the situations has been changed at runtime. However, the main shortcoming is the learning that may converge very slowly \cite{Jamshidi-fql}. The learning time can be shortened if the online learning entangled with offline training \cite{Bu2009}. This can be even further improved if we discover the relationship between parameters (\emph{e.g.}, \cite{Zheng2007,chen2009experience}) and exploit such knowledge at runtime.

\emph{Knowledge transfer}: There exist some approaches that reduce the configuration space by exploiting some knowledge about configuration parameters. Approaches like \cite{chen2009experience} use the dependence between the parameters in one system to facilitate finding optimal configuration in other systems. They embed the experience in a well-defined structure like Bayesian network through which the generation of new experiments can be guided toward the optimal region in other systems. 

\emph{Concluding remarks}: Software and systems community is not the only community that has tackled such problem. For instance, there exists interesting theoretical methods, \emph{e.g.} best arm identification problem for multi-armed bandit \cite{bubeck2011x}, that has been applied for optimizing hyper-parameters of machine learning algorithms, \emph{e.g.} supervised learning \cite{jamieson2015non}. More sophisticated methods based on surrogate models and meta-learning have reported better results in different areas, \emph{e.g.}, in propositional satisfiability problem \cite{hutter2011sequential}, convolutional neural networks \cite{snoek2012practical}, vision architectures \cite{bergstra2013making}, and more recently in deep neural networks \cite{domhan2015speeding}.

\begin{table}
			\caption{Systems configuration (auto-tuning) approaches.}
	\resizebox{\columnwidth}{!}{%
	\begin{threeparttable}
		\begin{tabular}{lcccl}
			\toprule
			Category & Empirical & Black box & Interactions & Approaches \\
			\midrule
			Rule-based  & No & No & No & \cite{db2advisor} \\
			DoE  & Yes & Yes & Yes & \cite{ustinova2015modelling}\\
			Model (white-box) & Partially & No & No & \cite{menasce2001preserving} \\
			Model (blackbox)  & Yes & Yes & Yes &\cite{johnstonperformance,yigitbasi2013towards,Siegmund}\\
			Search (sequential)  & Yes & Yes & Yes &\cite{xi2004smart,osogami2007optimizing,guo2010evaluating}\\
			Search (evol.)  & Partially & Yes & Yes & \cite{hansen2001completely,behzad2013taming,ye2003recursive}\\
			Space reduction & Yes & Yes & Yes & \cite{Zheng2007}\\
			Online learning & Yes & Yes & No & \cite{Bu2009}\\
			Knowledge transfer  & Yes & Yes & Yes & \cite{chen2009experience}\\
			\bottomrule
		\end{tabular}
		\begin{tablenotes}
			\small
			\item Empirical column describes whether the configuration is based on real data. Interactions describes whether the non-linear interactions can be supported. 
		\end{tablenotes}
	\end{threeparttable}}
	
	\label{tab:category-of-approaches}
\end{table}

%% file: sections/conclusions.tex
\section{Conclusions}
\label{sec:conclusions}

This paper proposes {\small \sf BO4CO}, an approach for locating optimal configurations using ideas of carefully choosing where to sample by sequentially reducing uncertainty in the response surface approximation in order to reduce the number of measurements. {\small \sf BO4CO} sequentially gains knowledge about the posterior distribution of the minimizers. We experimentally demonstrate that {\small\sf BO4CO} is able to locate the minimum of some benchmark functions as well as optimal configurations within real stream datasets accurately compared to five baseline approaches. We have carried out extensive experiments with three different stream benchmarks running on Apache Storm. The experimental results demonstrate that {\small \sf BO4CO} outperforms the baselines in terms of distance to the optimum performance with at least an order of magnitude. We have also provided some evidence that the learned model throughout the search process can be also useful for performance predictions. As a future work, since in the DevOps context several versions of a system are continuously delivered, we will use the notion of knowledge transfer \cite{kurekknowledge,chen2009experience} to accelerate the configuration tuning of the current version under test. 


%% file: sections/appendix.tex
\section{Storm Configuration Parameters}
\label{sec:appendix}

In this extra material, we briefly describe additional details about the experimental setting and complementary results that were not included in the main text. 

\section{Further Details on Settings}

\subsection{Code and Data}

{\noindent \small \url{https://github.com/dice-project/DICE-Configuration-BO4CO}}

\subsection{Documents}

{\noindent \small \url{https://github.com/dice-project/DICE-Configuration-BO4CO/wiki}}

\subsection{Configuration Parameters}

The list of configuration parameters in Apache Storm that we have used in the experiments (cf. Table \ref{tab:configuration-parameters-appendix}): 
	\begin{itemize}
		\setlength\itemsep{0em}
		\item {\sf \small max\_spout (topology.max.spout.pending)}. The maximum number of tuples that can be pending on a spout.
		\item {\sf \small spout\_wait (topology.sleep.spout.wait.strategy.time.ms)}. Time in ms the {\sf \small SleepEmptyEmitStrategy} should sleep.
		\item  {\sf \small netty\_min\_wait (storm.messaging.netty.min\_wait\_ms)}. The min time {\sf \small netty} waits to get the control back from OS. 
		\item {\sf \small spouts, splitters, counters, bolts}. Parallelism level. 
		\item {\sf \small heap}. The size of the worker heap.
		\item  {\sf \small buffer\_size (storm.messaging.netty.buffer\_size)}. The size of the transfer queue.
		\item {\sf \small emit\_freq (topology.tick.tuple.freq.secs)}. The frequency at which tick tuples are received.
		\item {\sf \small top\_level}. The length of a linear topology. 
		\item {\sf \small message\_size, chunk\_size}. The size of tuples and chunk of messages sent across PEs respectively.
	\end{itemize}

\subsection{Benchmark Settings}

Table \ref{tab:cluster-specification} represent the infrastructure specification we have used in the experiments (cf. testbed column in Table \ref{tab:configuration-parameters-appendix}).

\begin{table}[h!]
	\centering
	\caption{Cluster specification}
	\label{tab:cluster-specification}
	\resizebox{\columnwidth}{!}{%
		\begin{tabular}{@{}cl@{}}
			\toprule
			\textbf{Cluster} & \multicolumn{1}{c}{\textbf{Specification}}                                                                                   \\ \midrule
			C1               & \begin{tabular}[c]{@{}l@{}}OpenNebula, 3 Sup, 1 ZK, 1 Nimbus, N: (1CPU, 4GB Mem)\end{tabular}                               \\ \midrule
			C2               & \begin{tabular}[c]{@{}l@{}}EC2, 3 Sup, 1 ZK, 1 Nimbus, N: m1.medium (1 CPU, 3.75GB)\end{tabular}                          \\ \midrule
			C3               & \begin{tabular}[c]{@{}l@{}}OpenNebula, 3 Sup: (3CPU,6GB Mem), 1 ZK: (1CPU,4GB Mem),\\ 1 Nimbus: (2CPU,4GB Mem)\end{tabular} \\ \midrule
			C4               & \begin{tabular}[c]{@{}l@{}}EC2, 3 Sup, 1 ZK, 1 Nimbus, N: m3.large (2CPU, 7.5GB)\end{tabular}                             \\ \midrule
			C5               & \begin{tabular}[c]{@{}l@{}}Azure, 3 Sup: Standard\_D1(1CPU, 3.5GB) , 1 ZK, 1 Nimbus, \\ N: Standard\_A1(1CPU, 1.75GB) \end{tabular}                         \\ 
			\bottomrule
		\end{tabular}
	}
\end{table}
	
\subsection{Datasets}

Note that the parameters with $\star$ shows the interacting parameters. After collecting experimental data, we have used a common correlation-based feature selector implemented in Weka to rank parameter subsets according to a correlation based on a heuristic. The analysis results demonstrate that in all the 10 experiments at most 2-3 parameters were strongly interacting with each other, out of a maximum of 6 parameters varied simultaneously. Therefore, the determination of the regions where performance is optimal will likely to be controlled by such dominant factors, even though the determination of a global optimum will still depend on all the parameters.

\begin{table}[h!]
	\centering
	\caption{Experimental datasets, note that this is the complete set of datasets that we experimentally collected over the course of 3 months (24/7) for evaluating {\sf \small BO4CO}.} 
	\label{tab:configuration-parameters-appendix}
	\resizebox{\columnwidth}{!}{%
		\begin{threeparttable}
			\begin{tabular}{@{}lllgc@{}}
				\toprule
				&\textbf{Dataset} & \multicolumn{1}{c}{\textbf{Parameters}}                                                                                       & \multicolumn{1}{c}{\textbf{Size}} & \multicolumn{1}{c}{\textbf{Testbed}}                                                                             \\ \midrule
				1  & {\sf wc(6D) }               & \begin{tabular}[c]{@{}l@{}}{\sf $\star$1-spouts: \textcolor{blue}{\{1,3\}},}\\ {\sf $\star$2-max\_spout: \textcolor{blue}{\{1,2,10,100,1000,10000\}}, }\\ {\sf 3-spout\_wait:\textcolor{blue}{\{1,2,3,10,100\},}}\\ {\sf 4-splitters:\textcolor{blue}{\{1,2,3,6\}},}\\ {\sf $\star$5-counters:\textcolor{blue}{\{1,3,6,12\}},} \\ {\sf 6-netty\_min\_wait:\textcolor{blue}{\{10,100,1000\}}} \end{tabular}                      & 2880 & C1                 \\ \midrule
				2  & {\sf sol(6D) }             & \begin{tabular}[c]{@{}l@{}}{\sf $\star$1-spouts:\textcolor{blue}{\{1,3\}},} \\ {\sf $\star$2-max\_spout:\textcolor{blue}{\{1,10,100,1000,10000\}},} \\ {\sf $\star$3-top\_level:\textcolor{blue}{\{2,3,4,5\}},} \\ {\sf 4-netty\_min\_wait:\textcolor{blue}{\{10,100,1000\}},} \\ {\sf 5-message\_size: \textcolor{blue}{\{10,100,1e3,1e4,1e5,1e6\}},} \\ {\sf 6-bolts: \textcolor{blue}{\{1,2,3,6\}}} \end{tabular}                          & 2866 & C2                          \\ \midrule
				3  & {\sf rs(6D) }           & \begin{tabular}[c]{@{}l@{}}{\sf 1-spouts:\textcolor{blue}{\{1,3\}},} \\ {\sf 2-max\_spout:\textcolor{blue}{\{10,100,1000,10000\}}, }\\ {\sf $\star$3-sorters:\textcolor{blue}{\{1,2,3,6,9,12,15,18\}},} \\ {\sf 4-emit\_freq:\textcolor{blue}{\{1,10,60,120,300\}},}\\ {\sf 5-chunk\_size:\textcolor{blue}{\{1e5,1e6,2e6,1e7\}},} \\ {\sf 6-message\_size:\textcolor{blue}{\{1e3,1e4,1e5\}}} \end{tabular}                                    & 3840 & C3              \\ \midrule
				4  & {\sf wc(3D)}            & \begin{tabular}[c]{@{}l@{}}{\sf $\star$1-max\_spout:\textcolor{blue}{\{1,10,100,1e3, 1e4,1e5,1e6\}},} \\ {\sf $\star$2-splitters:\textcolor{blue}{\{1,2,3,4,5,6\}},} \\ {\sf 3-counters:\textcolor{blue}{\{1,2,3,4,5,6,7,8,9,10,11,12,13,14,15,16,17,18\}}} \end{tabular}                                                                             & 756 & C4                            \\ \midrule
				5  & {\sf wc+rs }          & \begin{tabular}[c]{@{}l@{}}{\sf $\star$1-max\_spout:\textcolor{blue}{\{1,10,100,1e3, 1e4,1e5,1e6\}},} \\ {\sf $\star$2-splitters:\textcolor{blue}{\{1,2,3,6\}},} \\ {\sf 3-counters:\textcolor{blue}{\{1,3,6,9,12,15,18\}}} \end{tabular}                                                                              & 196 & C4                                \\ \midrule
				6  & {\sf wc+sol }         & \begin{tabular}[c]{@{}l@{}}{\sf $\star$1-max\_spout:\textcolor{blue}{\{1,10,100,1e3, 1e4,1e5,1e6\}},} \\ {\sf $\star$2-splitters:\textcolor{blue}{\{1,2,3,6\}},} \\ {\sf 3-counters:\textcolor{blue}{\{1,3,6,9,12,15,18\}}} \end{tabular}                                                                                  & 196 & C4                              \\ \midrule
				7  & {\sf wc+wc }        & \begin{tabular}[c]{@{}l@{}}{\sf $\star$1-max\_spout:\textcolor{blue}{\{1,10,100,1e3, 1e4,1e5,1e6\}},} \\ {\sf $\star$2-splitters:\textcolor{blue}{\{1,2,3,6\}},} \\ {\sf 3-counters:\textcolor{blue}{\{1,3,6,9,12,15,18\}}} \end{tabular}                                                                       & 196 & C4                                \\ \midrule
				8  & {\sf wc(5D)}            & \begin{tabular}[c]{@{}l@{}}{\sf $\star$1-spouts:\textcolor{blue}{\{1,2,3\}},} \\ {\sf 2-splitters:\textcolor{blue}{\{1,2,3,6\}},} \\ {\sf 3-counters:\textcolor{blue}{\{1,2,3,6,9,12\}},} \\ {\sf 4-buffer-size:\textcolor{blue}{\{256k,1m,5m,10m,100m\}},} \\ {\sf 5-heap:\textcolor{blue}{\{``-Xmx512m", ``-Xmx1024m", ``-Xmx2048m"\}}} \end{tabular}                                                                     & 1080 & C5                            \\ \midrule
				9  & {\sf wc-c1}        & \begin{tabular}[c]{@{}l@{}}{\sf $\star$1-spout\_wait:\textcolor{blue}{\{1,2,3,4,5,6,7,8,9,10,100,1e3,1e4\}},} \\ {\sf $\star$2-splitters:\textcolor{blue}{\{1,2,3,4,5,6\}},}\\ {\sf 3-counters:\textcolor{blue}{\{1,2,3,4,5,6,7,8,9,10,11,12,13,14,15,16,17,18\}}} \end{tabular}                                                                               & 1343 & C1                          \\ \midrule
				10 & {\sf wc-c3}      & \begin{tabular}[c]{@{}l@{}}{\sf $\star$1-spout\_wait:\textcolor{blue}{\{1,2,3,4,5,6,7,8,9,10,100,1e3,1e4,6e4\}},} \\ {\sf $\star$2-splitters:\textcolor{blue}{\{1,2,3,4,5,6\}},}\\ {\sf 3-counters:\textcolor{blue}{\{1,2,3,4,5,6,7,8,9,10,11,12,13,14,15,16,17,18\}}} \end{tabular}                                                              & 1512 & C3 \\ 
				\bottomrule
			\end{tabular}
		\end{threeparttable}}
	\end{table}

\subsection{Performance gain}
The performance gain between the worst and best configuration settings are measured for each datasets in Table \ref{tab:performance-gain}.

\begin{table}[b]
	\centering
	\caption{Performance gain between best and worst settings.} 
	\label{tab:performance-gain}
	\resizebox{0.9\columnwidth}{!}{%
		\begin{threeparttable}
			\begin{tabular}{llccg}
				\toprule
				&\textbf{Dataset} & \multicolumn{1}{c}{\textbf{Best(ms)}}                                                                                       & \multicolumn{1}{c}{\textbf{Worst(ms)}} & \multicolumn{1}{c}{\textbf{Gain (\%)}}                                                                             \\ \midrule
				1  & {\sf wc(6D) }               &       55209          & 3.3172 & 99\%                 \\ \midrule
				2  & {\sf sol(6D) }         &           40499         & 1.2000 & 100\%                  \\ \midrule
				3  & {\sf rs(6D) }           &                34733           & 1.9000 & 99\%              \\ \midrule
				4  & {\sf wc(3D)}            &        94553           & 1.2994 & 100\%            \\ \midrule
				5  & {\sf wc(5D)}   &       405.5             & 47.387 & 88\%              \\ 
				
				\bottomrule
			\end{tabular}
		\end{threeparttable}}
	\end{table}

\subsection{How we set $\kappa$}
We set the exploration-exploitation parameter $\large \kappa$ (cf. Figure \ref{fig:kappa} and {\sf \small boLCB.m} in the github repository)) as:
\begin{align} \label{eq:kappa}
\kappa_t=\sqrt{2\log(|\mathbb{X}|\zeta(r)t^r/\epsilon)}, 
\zeta(r)=\sum_{n=1}^{\infty}\frac{1}{n^r}
\end{align}
where $0<\epsilon<1$ and $r\in \mathbb{N}, r\ge2$, $\zeta(r)$ is Riemann zeta.